\documentclass[twocolumn]{aastex631}
\usepackage{romannum}
\usepackage{rotating}

\newcommand{{\kms}}{{\mathrm{km\,s^{-1}}}}
\newcommand{{\Teff}}{{T_\mathrm{eff}}}
\newcommand{{\TLDR}}{{T_\mathrm{LDR}}}
\newcommand{{\FeH}}{{\rm {[Fe/H]}}}
\newcommand{{\XH}}[1]{{\rm {[{#1}/H]}}}
\newcommand{{\XFe}}[1]{{\rm {[{#1}/Fe]}}}
\newcommand{{\logP}}{{\log P}}
\newcommand{{\loggf}}{{\log gf}}
\newcommand{{\logg}}{{\log g}}
\newcommand{{\loggLDR}}{{\log g_\mathrm{LDR}}}
\newcommand{{\loggtrend}}{{\log g_\mathrm{trend}}}

\usepackage{hyperref}

\usepackage{tabularx}
\usepackage{amsmath}
\usepackage{rotating}

\def\testbx{bx}%
\DeclareRobustCommand{\ion}[2]{%
\relax\ifmmode
\ifx\testbx\f@series
{\mathbf{#1\,\mathsc{#2}}}\else
{\mathrm{#1\,\mathsc{#2}}}\fi
\else\textup{#1\,{\mdseries\textsc{#2}}}%
\fi}

\shorttitle{AASTeX v6.3.1 Sample article}
\shortauthors{Elgueta et al.}
\graphicspath{{./}{figures/}}
\usepackage{float}
\begin{document}
\pagenumbering{arabic}
\title{Astrophysical calibration of the oscillator strengths of YJ-band absorption lines in classical Cepheids}

\author[0000-0001-5642-2569]{Elgueta, S. S.}
\affiliation{Department of Astronomy, School of Science, The University of Tokyo, 7-3-1, Hongo, Bunkyo-ku, Tokyo 113-0033, Japan}
\affiliation{Instituto de Estudios Astrof\'isicos, Universidad Diego Portales, Av. Ej\'ercito Libertador 441, Santiago, Chile.}
\affiliation{Instituto de Astrof\'isica, Pontificia Universidad Cat\'olica de Chile, Av. Vicu\~na Mackenna 4860, 782-0436 Macul, Santiago, Chile.}
\affiliation{Millenium Nucleus ERIS, Instituto de Estudios Astrof\'isicos, Universidad Diego Portales, Av. Ej\'ercito Libertador 441, Santiago, Chile.}
\affiliation{Millennium Institute of Astrophysics, Av. Vicu\~na Mackenna 4860, 782-0436 Macul, Santiago, Chile.}

\author{Matsunaga, N.}
\affiliation{Department of Astronomy, School of Science, The University of Tokyo, 7-3-1, Hongo, Bunkyo-ku, Tokyo 113-0033, Japan}
\affiliation{Laboratory of Infrared High-Resolution spectroscopy (LiH), Kyoto Sangyo University, Kyoto, Japan.}

\author[0000-0002-5649-7461]{Jian, M.}
\affiliation{Department of Astronomy, School of Science, The University of Tokyo, 7-3-1, Hongo, Bunkyo-ku, Tokyo 113-0033, Japan}
\affiliation{Department of Astronomy
Stockholm University, AlbaNova University Center,
Roslagstullsbacken 21, 
114 21 Stockholm, Sweden}

\author[0000-0002-2861-4069]{Taniguchi, D.}
\affiliation{Department of Astronomy, School of Science, The University of Tokyo, 7-3-1, Hongo, Bunkyo-ku, Tokyo 113-0033, Japan}
\affiliation{National Astronomical Observatory of Japan, 2-21-1 Osawa, Mitaka,
Tokyo 181–8588, Japan.}

\author{Kobayashi, N.}
\affiliation{Kiso Observatory, Institute of Astronomy, School of Science, The University of Tokyo, Nagano, Japan.}

\author{Fukue, K.}
\affiliation{Laboratory of Infrared High-Resolution spectroscopy (LiH), Kyoto Sangyo University, Kyoto, Japan.}
\affiliation{Education Center for Medicine and Nursing, Shiga University of Medical Science, Seta Tsukinowa-cho, Otsu, Shiga 520-2192, Japan}

\author[0000-0002-6505-3395]{Hamano, S.}
\affiliation{Laboratory of Infrared High-Resolution spectroscopy (LiH), Kyoto Sangyo University, Kyoto, Japan.}
\affiliation{National Astronomical Observatory of Japan, 2-21-1 Osawa, Mitaka,
Tokyo 181–8588, Japan.}

\author[0000-0001-6401-723X]{Sameshima, H.}
\affiliation{Laboratory of Infrared High-Resolution spectroscopy (LiH), Kyoto Sangyo University, Kyoto, Japan.}
\affiliation{Institute of Astronomy, School of Science,
The University of Tokyo, Mitaka, Tokyo, Japan.}

\author{Kondo, S.}
\affiliation{Kiso Observatory, Institute of Astronomy, School of Science, The University of Tokyo, Nagano, Japan.}
\affiliation{Laboratory of Infrared High-Resolution spectroscopy (LiH), Kyoto Sangyo University, Kyoto, Japan.}

\author[0000-0002-5756-067X]{Arai, A.}
\affiliation{Laboratory of Infrared High-Resolution spectroscopy (LiH), Kyoto Sangyo University, Kyoto, Japan.}
\affiliation{Subaru Telescope, National Astronomical Observatory of Japan, 650
North Aohoku Place, Hilo, Hawaii, 96720, USA}

\author[0000-0003-2380-8582]{Ikeda, Y.}
\affiliation{Laboratory of Infrared High-Resolution spectroscopy (LiH), Kyoto Sangyo University, Kyoto, Japan.}
\affiliation{Photocoding, 460-102 Iwakura-Nakamachi, Sakyo-ku, Kyoto, 606–0025, Japan.}

\author[0000-0003-2011-9159]{Kawakita, H.}
\affiliation{Laboratory of Infrared High-Resolution spectroscopy (LiH), Kyoto Sangyo University, Kyoto, Japan.}
\affiliation{Department of Physics, Faculty of Sciences, Kyoto Sangyo University, Kyoto, Japan.}
\author{Otsubo, S.}
\affiliation{Laboratory of Infrared High-Resolution spectroscopy (LiH), Kyoto Sangyo University, Kyoto, Japan.}

\author{Sarugaku, Y.}
\affiliation{Laboratory of Infrared High-Resolution spectroscopy (LiH), Kyoto Sangyo University, Kyoto, Japan.}

\author[0000-0003-3579-7454]{Yasui, C.}
\affiliation{Laboratory of Infrared High-Resolution spectroscopy (LiH), Kyoto Sangyo University, Kyoto, Japan.}
\affiliation{National Astronomical Observatory of Japan, 2-21-1 Osawa, Mitaka,
Tokyo 181–8588, Japan.}

\author[0000-0002-9397-3658]{Tsujimoto, T.}
\affiliation{National Astronomical Observatory of Japan, 2-21-1 Osawa, Mitaka,
Tokyo 181–8588, Japan.}



\begin{abstract}
Newly-developed spectrographs with increased resolving powers, particularly those covering the near-IR range, allow the characterization of more and more absorption lines in stellar spectra. This includes the identification and confirmation of absorption lines and the calibration of oscillator strengths. In this study, we provide empirical values of $\loggf$ based on abundances of classical Cepheids obtained with optical spectra in \cite{Luck-2018}, in order to establish the consistency between optical and infrared abundance results. Using time-series spectra of classical Cepheids obtained with WINERED spectrograph (0.97--1.35\,$\mu$m, R $\sim$ 28000), we demonstrate that we can determine the stellar parameters of the observed Cepheids, including effective temperature ($\Teff$), surface gravity ($\logg$), microturbulence ($\xi$), and metallicity ($\XH{M}$). With the newly calibrated relations of line-depth ratios (LDRs), we can achieve accuracy and precision comparable to optical studies \citep{Luck-2018}, with uncertainties of $\sim$ 90\,K and 0.108\,dex for $\Teff$, and $\logg$, respectively. Finally, we created a new atlas of absorption lines, featuring precise abundance measurements of various elements found in the atmosphere of Cepheids (including neutron-capture elements), with $\loggf$ values that have been astrophysically calibrated.
\end{abstract}



\section{Introduction} \label{sec:intro}

Infrared spectroscopic observations provide valuable means for studying astronomical targets that are heavily obscured by interstellar extinction, which is a significant advantage over optical observations. The availability of mid-to-high resolution infrared spectra has led to the detection of both strong and weak absorption lines, The NIR regime, in particular, the \textit{YJ} bands, is a relatively understudied region that has the potential to yield significant new insights into the chemical evolution of the Galaxy. 

Despite the importance of basic spectral information, such as line positions and strengths in understanding the overall spectral scenario, the atomic information relevant to each spectral feature has not been widely covered in the literature \citep{Ryde-2019}. The calibration of these parameters is crucial for further procedures, such as estimating abundances. The oscillator strength ($\loggf$) is a dimensionless quantity that measures the strength of a transition (absorption and emission). In the NIR regime, accurately estimated $\loggf$ values are often lacking \citep[see, e.g.,][]{Andreasen-2016}, and proper line identification is also frequently absent. 

It is important to note that the definitive atomic parameters that are currently missing in the infrared bands must ultimately be obtained through laboratory astrophysics. However, in order to ensure consistency between optical and infrared chemical abundance results, we aim to obtain the empirical (or astrophysical) $\loggf$ values based on the optical abundance results for Cepheids presented in \citet{Luck-2018}. Such empirical calibrations also help to reduce the internal errors caused by the choice of the atmosphere model and, ultimately, improve our abundance measurements. The sample of \citet{Luck-2018} constitutes the largest collection of Cepheids that have been homogeneously analyzed for different elements in the optical range, making it an ideal data set for this purpose.

Cepheid variables are not only valuable for distance estimates, but their spectra also exhibit a considerable number of absorption lines, making them useful tracers of chemical abundances in their host galaxies, particularly in the case of the Galaxy. In particular, Cepheids have been proven to be useful for studying the metallicity gradient of the Galactic disk e.g., \citep{Genovali-2014, Luck-2018}. 
 A key parameter in the spectroscopic analysis is the effective temperature, $\Teff$. Once this intrinsic parameter is derived, the subsequent (but not less important) parameters, i.e., surface gravity ($\logg$), microturbulent velocity ($\xi$), and metallicity ($\FeH$) may be estimated with spectra. Besides the identification and calibration of absorption lines in Cepheids, we aim to exploit the capabilities of the unique high-resolution spectra with WINERED \citep{Ikeda-2022}, by employing the empirical LDR method \citep{Gray-1991}. We follow the prescription made in the NIR by \cite{Matsunaga-2021} but extend it for the first time in the $\Teff$ and $\logg$ parameter space to cover the range of classical Cepheids.

This paper is organized as follows. In Section~\ref{sec:data}, we provide detailed information about the analyzed targets and the WINERED spectrograph, as well as the procedure employed to prepare the spectra for measurements. In Section~\ref{sec:stellar_param}, we discuss the capabilities of our data to provide precise and accurate stellar parameters ($\Teff$ and $\logg$, primarily) based on our extension of the LDR method. Section~\ref{sec:lines} presents our method to identify and confirm the lines we will use for abundance determination, as well as the derivation of microturbulence ($\xi$) and the calibration of the oscillator strengths ($\loggf$). Based on the results in Section~\ref{sec:lines}, we measured the abundances of Cepheids as we present in Section~\ref{sec:ab_analysis}. We later discuss the scope and limitations of our results in
 Section~\ref{sec:conclussion}.

\section{Data} \label{sec:data}
\subsection{Observations and targets}
We analyze multi-epoch spectra of eleven Cepheids in the solar neighborhood obtained with the WINERED spectrograph in its WIDE mode, which has a resolving power of R $\sim$ 28000. The observations were conducted from 2013 to 2016 at the 1.3 m Araki Telescope located at the Koyama Observatory of Kyoto Sangyo University in Japan. As given in Tables~\ref{tab:obs_log_calibrators} and \ref{tab:obs_log_val}, our sample is divided into two groups, calibrators, and validators. The former consists of eight widely studied Cepheids for which there is a wealth of data sets available (photometric and spectroscopic). The latter consists of three Cepheids whose information in the literature is incomplete or scarce. For each target, we obtained multi-epoch spectra, with better phase coverage for the calibrators.

\subsection{Pipeline reduction}
During the observations, the nodding technique was employed to subtract the background radiation from the sky and the ambient facility. Two types of nodding patterns were used, ``ABBA'' and ``OSO'' (object-sky-object). We decided the nodding pattern for each observation to avoid latent signals after exposure of bright targets. Calibration data include flat fielding images and ThAr lamp spectra for wavelength calibration. 
 The raw data were reduced by the WARP pipeline, created and developed by Hamano, S. \footnote{\href{https://github.com/SatoshiHamano/WARP}{WARP in Github}}. Every science target has more than a single exposure, and for different echelle orders, a combined 1D, air wavelength calibrated spectrum was obtained. The pipeline also outputs information about the signal-to-noise ratio (S/N) for each target at three different regions of each echelle order. A WINERED spectrum in the WIDE mode is divided into 20 echelle orders, of which we use the 11 orders covering  9760--10890\,{\AA} and 11800--13190\,{\AA} which are not heavily affected by telluric contamination.

{\small
\begin{longtable*}{lccccccc}
\caption{Observation Log (Calibrators)} \label{tab:obs_log_calibrators}\\
\hline\hline
\textbf{Target} & \textbf{RA} & \textbf{DEC} & \textbf{Date and Time (UT)} & \textbf{JD} & \textbf{Phase} & \textbf{Exposures} & \textbf{Nodding} \\ 
\hline\hline
\endfirsthead
\endlastfoot
\hline
  $\delta$ Cep &  22:29:10.27 & $+$58:24:54.7 & 2013-12-02 11:12 & 2456628.967 & 0.119 & $188^s \times 3$ & ABB \\
            &  & & 2013-12-05 10:24 & 2456631.934 & 0.672 & $188^s \times 2$ & AB  \\
         & & & 2013-12-25 10:07 & 2456651.922 & 0.396 & $138^s \times 4$ & OSO  \\
         & & & 2015-08-06 13:09 & 2457241.048 & 0.178 & $20^s \times 2$ & OSO  \\
         & & & 2015-08-08 13:49 & 2457243.076 & 0.556 & $20^s \times 4$ & ABBA \\
         & & & 2015-08-15 18:34 & 2457250.274 & 0.897 & $30^s \times 2$ & AB \\
         & & & 2015-10-23 12:51 & 2457319.036 & 0.711 & $30^s \times 4$ & OSO   \\
\hline
  $\eta$ Aql & 19:52:28.37 & $+$01:00:20.4 & 2014-09-16 13:00 & 2456917.042 & 0.743 & $188^s \times 2$ & AB \\
          &  & & 2014-09-17 10:23 & 2456917.933 & 0.867 & $18^s \times 4$ & ABBA \\
          &  & & 2015-08-06 12:33 & 2457241.023 & 0.887 & $30^s \times 4$ & ABBA  \\
          &  & & 2015-08-08 13:59 & 2457243.083 & 0.174 & $30^s \times 4$ & ABBA  \\
          &  & & 2016-03-21 19:58 & 2457469.332 & 0.700 & $100^s \times 2$ & OSO  \\
          &  & & 2016-03-26 19:46 & 2457474.324 & 0.395 & $30^s \times 6$ & OSO   \\
          &  & & 2016-05-14 18:36 & 2457523.275 & 0.216 & $50^s \times 2$ & OSO \\
\hline
  FF Aql & 18:58:14.75 & $+$17:21:39.3 & 2015-08-06 14:48 & 2457241.117 & 0.594 & $120^s \times 4$ & OSO \\
         & & & 2015-08-07 15:23 & 2457242.141 & 0.823 & $60^s \times 2$ & OSO  \\
         & & & 2016-03-17 19:13 & 2457465.301 & 0.737 & $150^s \times 4$ & OSO  \\
         & & & 2016-03-21 18:41 & 2457469.279 & 0.627 & $300^s \times 2$ & OSO \\
         & & & 2016-04-19 17:49 & 2457498.243 & 0.105 & $300^s \times 2$ & AB  \\
\hline
  RT Aur &  06:28:34.09 & $+$30:29:34.9 & 2014-01-23 16:20 & 2456681.181 & 0.920 & $228^s \times 4$ & ABBA \\
         & & & 2015-10-28 18:53 & 2457324.287 & 0.422 & $150^s \times 4$ & ABBA  \\
         & & & 2016-02-16 14:41 & 2457435.112 & 0.149 & $300^s \times 2$ & OSO  \\
         & & & 2016-02-28 12:20 & 2457447.014 & 0.341 & $300^s \times 2$ & OSO \\
         & & & 2016-03-07 12:47 & 2457455.033 & 0.492 & $150^s \times 2$ & OSO  \\
         & & & 2016-03-15 11:12 & 2457462.967 & 0.620 & $300^s \times 4$ & OSO \\
         & & & 2016-03-17 11:03 & 2457464.961 & 0.155 & $200^s \times 4$ & ABBA \\
         & & & 2016-03-21 13:13 & 2457469.051 & 0.252 & $200^s \times 4$ & OSO \\
         & & & 2016-03-23 12:05 & 2457471.004 & 0.776 & $300^s \times 2$ & AB \\
\hline
  SU Cas & 02:51:58.75 & $+$68:53:18.6 & 2013-12-25 10:07 & 2456651.922 & 0.054 & $588^s \times 4$ & ABBA \\
        & & & 2015-08-15 18:50 & 2457250.285 & 0.014 & $120^s \times 2$ & AB  \\
        & & & 2016-03-12 09:36 & 2457459.900 & 0.546 & $300^s \times 2$ & OSO  \\
        & & & 2016-03-15 10:40 & 2457462.945 & 0.108 & $300^s \times 2$ & OSO  \\
        & & & 2016-03-17 11:24 & 2457464.975 & 0.150 & $300^s \times 4$ & ABBA \\
\hline
  SZ Tau & 04:37:14.78 & $+$18:32:34.9 & 2014-01-24 13:00 & 2456682.042 & 0.870 & $288^s \times 4$ & ABBA  \\
        & & & 2016-02-18 12:12 & 2457437.009 & 0.639 & $300^s \times 4$ & ABBA \\
        & & & 2016-03-17 09:57 & 2457464.915 & 0.502 & $300^s \times 2$ & AB \\
        & & & 2016-03-21 09:59 & 2457468.916 & 0.772 & $300^s \times 2$ & OSO  \\
        & & & 2016-03-25 11:00 & 2457472.959 & 0.056 & $300^s \times 4$ & ABBA \\
\hline
  X Cyg & 20:43:24.19 & $+$35:35:16.1 & 2014-08-30 13:16 & 2456900.053 & 0.606 & $188^s \times 4$ & ABBA \\
        & & &  2014-09-15 13:03 & 2456916.044 & 0.582 & $108^s \times 4$ & ABBA \\
        & & & 2014-09-18 11:12 & 2456918.967 & 0.760 & $108^s \times 2$ & AB \\
        & & & 2014-10-15 11:13 & 2456945.968 & 0.408 & $138^s \times 2$ & AB \\
        & & & 2015-07-25 13:35 & 2457229.066 & 0.685 & $300^s \times 4$ & ABBA \\
        & & & 2015-10-26 10:49 & 2457321.951 & 0.353 & $300^s \times 2$ & OSO  \\
        & & & 2016-03-17 19:58 & 2457465.332 & 0.103 & $200^s \times 4$ & OSO \\
        & & & 2016-05-04 19:06 & 2457513.296 & 0.030 & $300^s \times 4$ & ABBA \\
        & & & 2016-05-18 17:16 & 2457527.220 & 0.880 & $300^s \times 5$ & OSO \\
\hline
  $\zeta$ Gem & 07:04:06.5 & $+$20:34:13.1 & 2013-02-22 11:38 & 2456345.985 & 0.534 & $188^s \times 7$ & OSO \\
          & & & 2013-02-23 14:35 & 2456347.108 & 0.645 & $108^s \times 8$ & OSO \\
          & & & 2013-02-27 12:01 & 2456351.001 & 0.028 & $108^s \times 9$ & OSO \\
          & & & 2013-03-03 14:38 & 2456355.110 & 0.433 & $88^s \times 9$ & OSO \\
          & & & 2013-11-29 15:08 & 2456626.131 & 0.133 & $108^s \times 4$ & ABBA \\
          & & & 2016-02-17 13:35 & 2457436.066 & 0.923 & $100^s \times 4$ & OSO \\
          & & & 2016-03-07 12:37 & 2457455.026 & 0.791 & $60^s \times 2$ & OSO  \\
          & & & 2016-03-23 11:13 & 2457470.968 & 0.362 & $180^s \times 4$ & OSO  \\
          & & & 2016-05-01 10:16 & 2457509.928 & 0.200 & $30^s \times 4$ & OSO \\
          & & & 2017-12-05 08:12 & 2458092.842 & 0.626 & $30^s \times 2$ & OSO \\
\hline
\end{longtable*}
}

{\small
\begin{longtable*}{lccccccc}
\caption{Observation Log (Validators)} \label{tab:obs_log_val}\\
\hline\hline
\textbf{Target} & \textbf{RA} & \textbf{DEC} & \textbf{Date and Time (UT)} & \textbf{JD} & \textbf{Phase} & \textbf{Exposures} & \textbf{Nodding} \\ 
\hline\hline
  DL Cas & 00:29:58.59 & $+$60:12:43.1 
 & 2015-07-31 17:36 & 2457235.234 & 0.743 & $300^s \times 4$ & ABBA  \\
        & & & 2015-08-07 14:49 & 2457242.118 & 0.604 & $600^s \times 2$ & OSO \\
        & & & 2015-10-23 13:07 & 2457319.047 & 0.219 & $300^s \times 4$ & OSO\\
\hline
  S Sge & 19:56:01.26 & $+$16:38:05.2  & 2015-10-26 10:29 & 2457321.937 & 0.940 & $300^s \times 2$ & OSO \\
        & & & 2016-03-21 19:01 & 2457469.293 & 0.520 & $300^s \times 2$ & OSO \\
        & & & 2016-03-26 19:36 & 2457474.317 & 0.119 & $300^s \times 2$ & AB \\
\hline
  T Vul & 20:51:28.24 & $+$28:15:01.8 & 2015-08-08 15:47 & 2457243.158 & 0.138 & $200^s \times 4$ & ABBA \\
        & & & 2016-05-14 18:48 & 2457523.284 & 0.293 & $150^s \times 2$ & OSO  \\
\hline
\end{longtable*}
}


\subsection{Telluric correction and continuum normalization}
The analysis of infrared spectra is subject to telluric contamination, caused by the Earth's atmosphere. The telluric lines vary depending on factors such as the observatory location, weather conditions, and air mass. In order to accurately identify and study stellar features, it is necessary to remove telluric contamination from the observed spectra. In this study, the method developed in \citet{Sameshima-2018} was employed to correct for the telluric contamination in the telluric standard stars observed in conjunction with the science targets. This correction was applied to every target and every order, with the exception of orders 53 and 54, corresponding to the spectral range covering $\lambda: [10280-10680]$ where the telluric absorption in our atmosphere can be neglected. The resulting spectra were then normalized by means of the IRAF continuum task to account for any offset from unity. 

\section{Stellar Parameters}
\label{sec:stellar_param}

\subsection{Phased parameters from literature data}
For the calibrators, we derived $\Teff$, $\logg$, and $\xi$ at each phase from the interpolated Fourier curves based on the catalog of \citet{Luck-2018}, which provides the measurement of these parameters with good phase sampling. We adopted the interpolated values for further analysis and calibrations. On the other hand, a different approach was applied to the validators by employing the Line-Depth Ratio (LDR) method.

\subsection{Construction of LDR relations}
\label{sec:LDR-method}

The LDR method is widely used for determining the effective temperature ($\Teff$) in spectroscopic analysis. Previous studies were primarily conducted on optical spectra \citep[][and references 
therein]{Gray-2001,Kovtyukh-2000,Kovtyukh-2007}, while more recent works have been involved with infrared spectra \citep{Fukue-2015,Taniguchi-2018,Jian-2019}. However, all of these previous studies were limited to certain temperature ranges which do not include warmer Cepheids ($\Teff > 6000$ K). Furthermore, LDRs used in the previous works tend to exhibit dependency not only on $\Teff$ but also on $\logg$ and metallicity \citep{Jian-2019,Jian-2020}. In order to bypass such complexities, \citet{Taniguchi-2021} proposed to use only \ion{Fe}{i} lines to estimate $\Teff$, followed by the work of \citet{Matsunaga-2021}, introducing the usage of neutral-ionized pairs to give the LDR relations with $\logg$, setting the guideline of this study.

In this work, the LDR method is employed following the prescription stated in \citet{Matsunaga-2021} by using \ion{Fe}{i}, \ion{Fe}{ii}, \ion{Ca}{i}, and \ion{Ca}{ii} within 9760--10860\,{\AA} and 11800--13190\,{\AA} corresponding to the WINERED echelle orders in the $Y$ band (52nd--57th) and the $J$ band (43rd--47th).  From \citet{Matsunaga-2021}, we adopted the list of 97 lines for the four species (76 \ion{Fe}{i}, 5 \ion{Fe}{ii}, 11 \ion{Ca}{i} and 5 \ion{Ca}{ii}), but we searched for new line pairs that show good LDR relations. According to our preliminary test on our sample of Cepheids, the LDR relations  presented in \citet{Matsunaga-2021} do not work well in hotter targets ($\Teff > 6000$ K) as some lines are too shallow at high temperatures. It is worth mentioning that other elements were found in the spectra such as Si and Ti displaying both neutral and ionized lines but the number of ionized lines is very limited.

In order to measure the depth of each of the 97 lines from \citet{Matsunaga-2021}, a Gaussian fit was performed on the section of five pixels around the line center ($\lambda_c$). The wavelength of the line minimum ($\lambda_0$) was then determined, and assuming that the continuum normalization was properly done, the depth was calculated as the ``distance" from the continuum level (the unity) to the flux at $\lambda_0$. We rejected measurements if the continuum level was not well normalized and the estimated depth was a negative number, or if the position of $\lambda_0$ was too far from $\lambda_c$. This procedure was carried out using the $ir\_ldr$ python package developed by Jian, M.  \footnote{\href{https://github.com/MingjieJian/ir_ldr}{$ir\_ldr$ in Github}}. The error in each depth measurement is linked to the signal-to-noise ratio (S/N) for the target (obj) and the telluric standard (tell), and the two errors were added in quadrature:
\begin{equation}
    e  = \sqrt{e^{2}_{obj} + e^{2}_{tell}} = \sqrt{\Bigg(\frac{1}{(S/N)_{obj}}\Bigg)^2 + \Bigg(\frac{1}{(S/N)_{tell}}\Bigg)^2} \label{eq:errindepth}
\end{equation}

This equation was used for all the WINERED echelle orders except for the orders 53 and 54, for which no telluric correction was performed, as they cover the wavelength ranges corresponding to an atmospheric window free of telluric absorption. For those two orders, the error is simply $e=e_{obj}$. We created a new list of line pairs that are effective in the Cepheids regime 5500 K $<$ $\Teff$ $<$ 6500 K, following the algorithm of pair selection in \citet[Section~3.3]{Matsunaga-2021}.

\begin{table*}
\center{
\caption{LDR relations\label{tab:LDR_relations}}
\footnotesize
\begin{tabularx}{\textwidth}{cccccccccc}
\cline{1-10}
ID & Line 1 & Line 2 & Form ID & Form (ext) & $\alpha$ & $\beta$ & $\gamma$ & $\sigma_{y}$ & $\sigma_{p}$ \\
\cline{1-10}
(1)  & FeI  10155.162 & FeI  10353.804 & T4  & 21.655   & $-5.781$ & ---                   & 0.0506 &  109 &  39 \\
(2)  & FeI  10167.468 & FeI  10347.965 & T4  & 21.630   & $-5.812$ & ---                   & 0.0587 &  126 &  39 \\
(3)  & FeI  10195.105 & FeI  10216.313 & T3  & 2.4913   & $-5.657\times 10^{-4}$ & ---                   & 0.0659 &  117 &  46 \\
(4)  & FeI  10218.408 & FeI  10227.994 & T2  & 74.224   & $-19.296$ & ---                   & 0.2475 &  159 &  43 \\
(5)  & FeI  10340.885 & FeI  10532.234 & T2  & 18.883   & $-4.816$ & ---                   & 0.0407 &  105 &  50 \\
(6)  & FeI  10395.794 & FeI  9861.7337 & T2  & 18.502   & $-4.697$ & ---                   & 0.0401 &  106 &  45 \\
(7)  & FeI  10423.743 & FeI  10863.518 & T2  & 22.176   & $-5.767$ & ---                   & 0.0472 &  102 &  37 \\
(8)  & FeI  10577.139 & FeI  10849.465 & T2  & 27.377   & $-7.163$ & ---                   & 0.0729 &  126 &  34 \\
(9)  & FeI  10616.721 & FeI  9868.1857 & T2  & 21.268   & $-5.594$ & ---                   & 0.0583 &  130 &  32 \\
(10) & FeI  10783.050 & FeI  9889.0351 & T2  & 17.597   & $-4.530$ & ---                   & 0.0465 &  128 &  44 \\
(11) & FeI  10818.274 & FeI  9811.5041 & T4  & 16.784   & $-4.478$ & ---                   & 0.0391 &  108 &  42 \\
(12) & FeI  12556.996 & FeI  12648.741 & T2  & 21.438   & $-5.559$ & ---                   & 0.0624 &  139 &  31 \\
(13) & FeI  9868.1857 & FeII 9997.5980 & TG4 & 26.300   & $-7.183$ & $0.2368               $ & 0.0506 & 0.21 &  74 \\
(14) & FeI  10347.965 & FeII 10173.515 & TG3 & 1.9293   & $-4.147\times 10^{-4}$ & $0.2456               $ & 0.0698 & 0.28 &  57 \\
(15) & FeI  10353.804 & FeII 10366.167 & TG4 & 37.969   & $-10.217$ & $0.2699               $ & 0.0750 & 0.28 &  60 \\
(16) & FeI  10611.686 & FeII 10501.500 & TG3 & 2.0781   & $-4.782\times 10^{-4}$ & $0.2261               $ & 0.0489 & 0.22 &  88 \\
(17) & FeI  10818.274 & FeII 10862.652 & TG3 & 3.7638   & $-7.925\times 10^{-4}$ & $0.2587               $ & 0.0639 & 0.25 &  61 \\
(18) & CaI  10838.970 & CaII 9854.7588 & TG2 & 19.853   & $-5.306$ & $0.2080               $ & 0.0666 & 0.32 &  66 \\
(19) & CaI  10343.819 & CaII 9890.6280 & TG2 & 44.573   & $-11.798$ & $0.3498               $ & 0.1147 & 0.33 &  75 \\
(20) & CaI  10846.792 & CaII 9931.3741 & TG2 & 7.5153   & $-2.013$ & $0.0829               $ & 0.0369 & 0.45 &  65 \\
(21) & CaI  12105.841 & CaII 11838.997 & TG2 & 5.2987   & $-1.427$ & $0.0657               $ & 0.0300 & 0.46 &  69 \\
(22) & CaI  13033.554 & CaII 11949.744 & TG2 & 11.193   & $-2.990$ & $0.1109               $ & 0.0328 & 0.30 &  71 \\

\cline{1-10}
\end{tabularx}}
\end{table*}

To begin with the construction of our LDR relations, we use \ion{Fe}{i}--\ion{Fe}{i} pairs for estimating $\Teff$, by pairing lines of low and high excitation potential (having a $\Delta\mathrm{EP} \geq 1$ eV) gives higher sensitivity to $\Teff$ and their ratio can provide good diagnostics on temperature, reaching the utmost precision of $\sim$ 10 K \citep{Gray-1991}. Thus, the LDR of each pair is defined as $r = d_\mathrm{low}/d_\mathrm{high}$, i.e., the ratio of the depth of the line with a lower excitation potential ($d_\mathrm{low}$) to the depth of the line with a higher excitation potential ($d_\mathrm{high}$). We consider the four forms of the relation between $r$ and $\Teff$ as described in \citet[Section~3.4.1]{Matsunaga-2021}. We have 57 spectra of the calibrators in total, but the quality of the spectra shrank our sample to 51 (listed in table ~\ref{tab:obs_log}), as well as the validation of measured depths may reduce the number of useful depths and according to the number of ratios.
We rejected the line pairs for which only 30 or fewer validated measurements of $r$ were available and those for which the LDR relations had dispersion larger than 200 K. In addition, the range of $\Teff$ covered by the points used for the fitting must be larger than 1000 K. This last condition is considered in order to make each LDR relation useful for a wide range of targets. For each line pair, one of the four forms (T1)--(T4) was selected to give the most negligible dispersion.

Surface gravity ($\logg$) is also key in deriving subsequent stellar parameters, and analogously to $\Teff$ there are multiple ways to estimate it. Its calculation is, however, not trivial. Some approaches rely on numerical models that are computationally expensive, and others depend on how stellar masses and radii can be estimated with photometry and interferometry. In order to estimate $\logg$, we consider pairs of neutral and ionized lines, i.e., \ion{Fe}{i}--\ion{Fe}{ii} and \ion{Ca}{i}--\ion{Ca}{ii} pairs taken from the line list of \citet{Matsunaga-2021}. Their LDRs are defined as
\vspace{0.5cm}

$r=d_{\mbox{\ion{Fe}{i}}}/d_{\mbox{\ion{Fe}{ii}}}$ or $r=d_{\mbox{\ion{Ca}{i}}}/d_{\mbox{\ion{Ca}{ii}}}$.

\vspace{0.5cm}

We selected the set of the LDR relations, among those which were not rejected, in order to include each line only in one line pair. We finally obtained 12 \ion{Fe}{i}--\ion{Fe}{i}, 5 \ion{Fe}{i}--\ion{Fe}{ii}, and 5 \ion{Ca}{i}--\ion{Ca}{ii} relations (Table~\ref{tab:LDR_relations}).

\subsection{Application of LDR relations}

Using the relations we obtained in Section ~\ref{sec:LDR-method}, we can determine $\Teff$ and $\logg$ as follows: \\
\noindent
(1)~First, the LDR for each of the \ion{Fe}{i}--\ion{Fe}{i} pairs can be converted
to an estimate of effective temperature, $T_{i}$. 
We also estimate its error, $e_i$, considering the error in the LDR and
the scatter of the LDR relation ($\sigma_p$ in Table~\ref{tab:LDR_relations}).
Then, we calculate the weighted average and its standard error by
\begin{eqnarray}
T_\mathrm{LDR} &=& \sum_{i=1}^{N_\mathrm{pair}} w_i T_{\mathrm{LDR}, i} \Big/ \sum_{i=1}^{N_\mathrm{pair}} w_i \label{eq:TLDR} \\
e_T &=& \sqrt{ \frac{\sum_{i=1}^{N_\mathrm{pair}} w_i (T_{\mathrm{LDR}, i}-T_\mathrm{LDR})^2}{ (N_\mathrm{pair}-1) \sum_{i=1}^{N_\mathrm{pair}} w_i} }
\label{eq:TLDR_error}
\end{eqnarray}
where $w_i$ is a weight given by $1/e_i^2$.
\\
\noindent
(2) We then estimate the LDR-based surface gravity by using the $\TLDR$ obtained above. Based on each LDR relation of a \ion{Fe}{i}--\ion{Fe}{ii} pair
or a \ion{Ca}{i}--\ion{Ca}{ii} pair, we obtain an estimate of surface gravity,
$\logg_i$, and its error. Then, we calculate
the weighted mean ($\loggLDR$) and the error according to the formulae similar to
Equations~(\ref{eq:TLDR}) and (\ref{eq:TLDR_error}).

\begin{figure*}[htbp]
  \centering
 
  \caption{Performance of our LDR method for $\Teff$ and $\logg$}
  \begin{minipage}[t]{0.49\linewidth}
    \includegraphics[width=\linewidth]{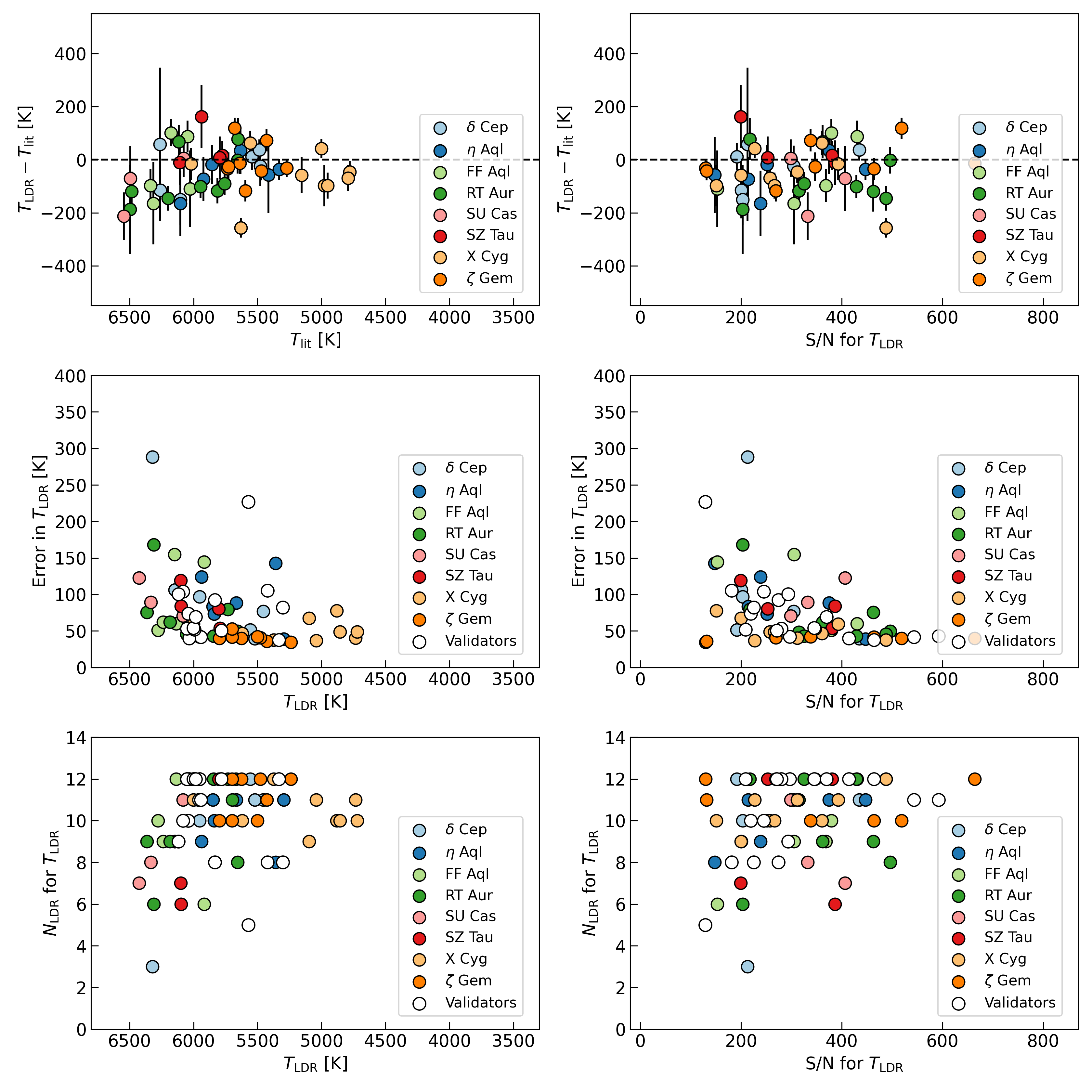}
    \label{fig:T_LDR_dev}
    
  \end{minipage}
  \hfill
  \begin{minipage}[t]{0.49\linewidth}
    \includegraphics[width=\linewidth]{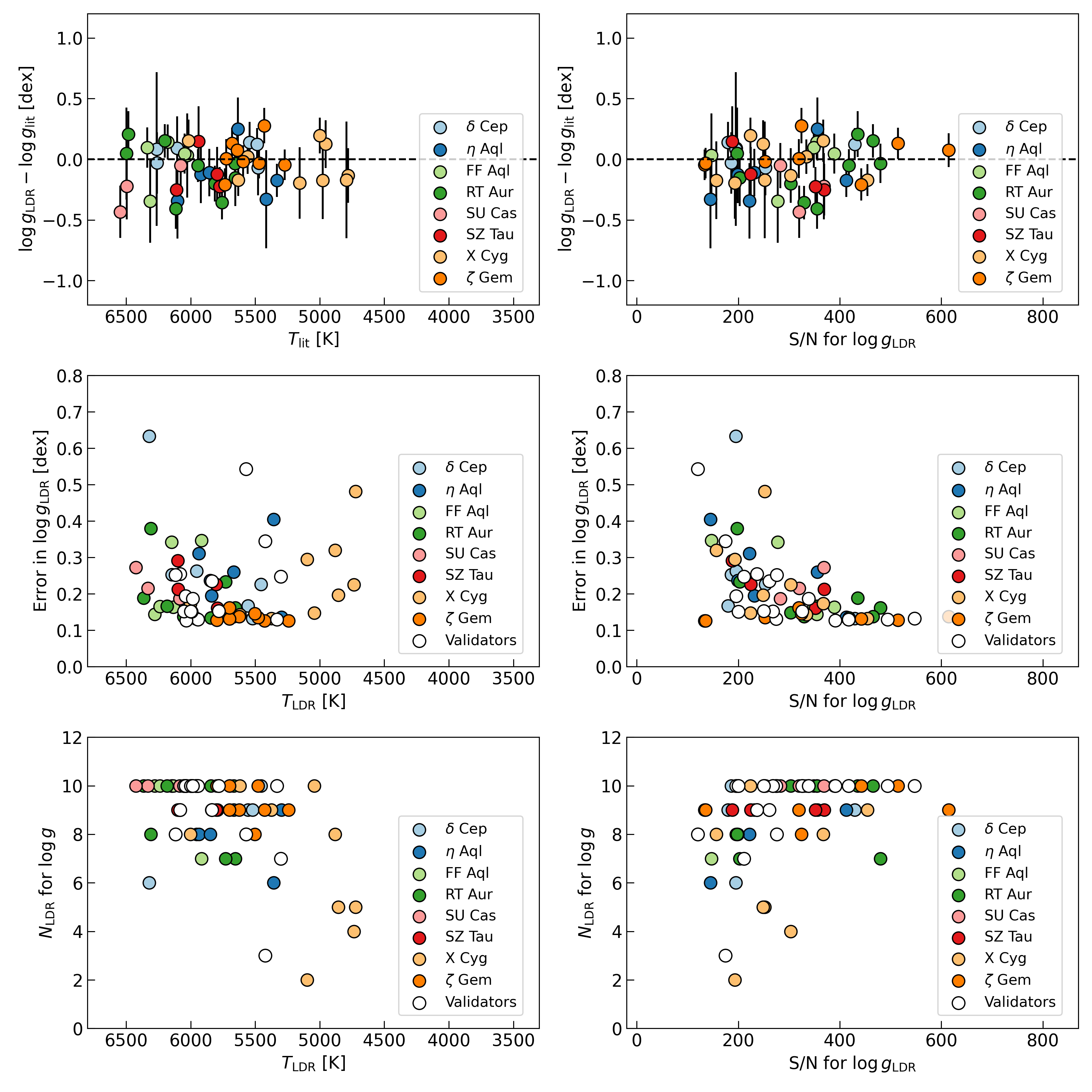}
    \label{fig:G_LDR_dev}
  \end{minipage}

  \label{fig:ldr_all}
\end{figure*}

Figure~\ref{fig:ldr_all} summarizes
the results for $\TLDR$ and $\loggLDR$ we obtained for the calibrators together with the results for the validators. The derived $\Teff$ and $\logg$ for each phase are listed in the supplementary material.
The deviations of $\TLDR$ and $\loggLDR$ from those expected from
the literature time-series data show the standard deviations
$\sim 90$\,K and $\sim 0.2$\,dex.
Larger errors are found at higher temperatures for $\TLDR$ or at both higher and lower temperatures
for $\loggLDR$.
Absorption lines of neutral atoms tend to get weak at higher temperatures,
while lines of ions tend to get weak at lower temperatures within
the range of interest. This trend is observed in 
the number of the line pairs in the bottom panels and explains
the larger deviations and errors seen in the middle panels.
The standard deviation of $\TLDR$, $\sim$ 90$\,K$, is larger than
42\,K found by \citet{Matsunaga-2021} for stars with $4800\leq \Teff \leq 6200$\,K
and $1.35\leq \logg\leq 4.5$, while the standard deviation of $\loggLDR \sim 0.2$\,dex, is similar to 0.17\,dex obtained in the aforementioned work.
The larger standard deviation of $\TLDR$ can be ascribed to
the fact that our sample includes significantly more stars with higher temperatures. The errors depend also on $S/N$. We will discuss this point in Section~\ref{sec:conclussion}.

\begin{figure}
\centering
\includegraphics[width=\linewidth]{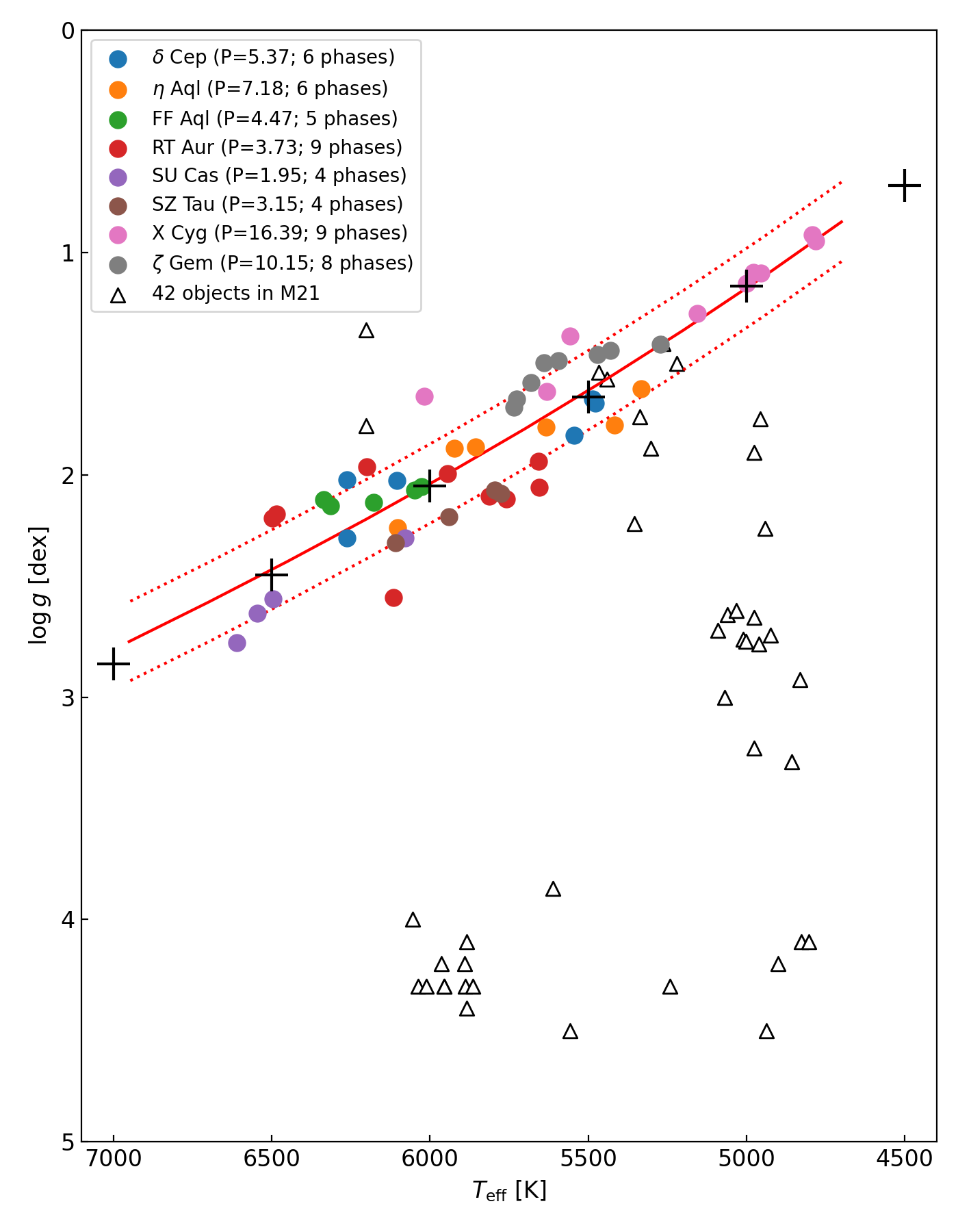}
\caption[Distribution of Cepheids' ($\Teff$, $\logg$)]{Distribution of the ($\Teff, \logg$) of the Calibrator Cepheids given in the literature \citep{Luck-2004,Luck-2008,Andrievsky-2005,Kovtyukh-2005} plotted altogether with the sample of 42 FGK stars (M21) from \citet{Matsunaga-2021}}
\label{fig:tefflogg}
\end{figure}

\subsection{Dependency of $\logg$ on $\Teff$ and P}
\label{teff-logg-logp}
Looking at the ($\Teff, \logg$) obtained with the literature curves, for the calibrators sample, it can be noted that the two parameters are tightly correlated (Figure~\ref{fig:tefflogg}). Such a correlation can be expected given the fact that the mean stellar parameters of Cepheids fall within the Cepheid instability strip, i.e., their mean $\Teff$, $\logg$, and also the luminosity $\textit{L}$ should be (anti-) correlated. However,  $\Teff$ at individual phases are not necessarily within the instability strip. Moreover, the variations of $\Teff$ and $\textit{L}$ do not follow the trend of the instability strip. A Cepheid becomes fainter when it gets cooler, while the instability strip makes fainter Cepheids warmer on average. Nevertheless, Figure~\ref{fig:tefflogg} shows a tight correlation between
$\Teff$ and $\logg$ at individual phases of the calibrator Cepheids.
Fitting the 51 points available for 8 calibrators in total,
we obtained the relation,
\begin{equation}
\logg = 11.110 \cdot \, \log (\Teff/5800) + 1.877
\end{equation}
with a scatter of 0.179\,dex. \\

\begin{figure*}
    \centering
    \includegraphics[width=\textwidth]{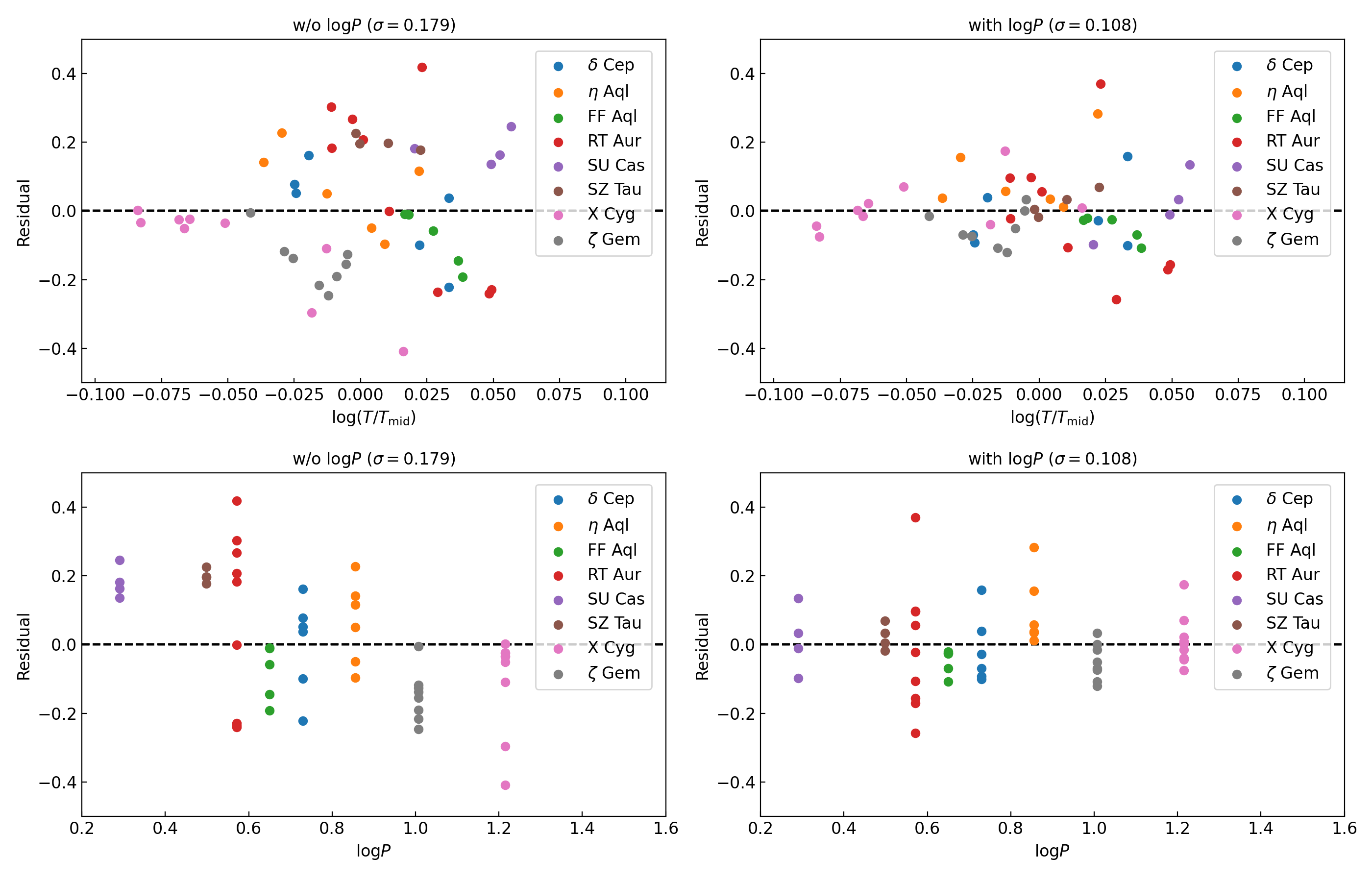}
    \caption[$\Teff$--$\logP$--$\logg$ trend of Cepheids]{$\Teff$--$\logP$--$\logg$ trend, displaying the dispersion when the pulsation period $P$ is included (right) or not (left) in the relation. $T_\mathrm{mid}=5800$.}
    \label{fig:tefflogPlogg}
\end{figure*}

Furthermore, we found that including the $\log P$ term reduces the scatter. As illustrated in Figure~\ref{fig:tefflogPlogg},
the residual around the $\Teff$--$\logg$ without the $\log P$ relation shows
the dependency on the period. With the $\log P$ term included,
we obtained the relation,
\begin{equation}
\logg = 6.483 \cdot \log (\Teff/5800) -0.775 \cdot \log P + 2.475
\label{eq:tefflogglogp}
\end{equation}
with the scatter of 0.108\,dex.
This is smaller than the errors in the $\loggLDR$ presented above. It is hard to estimate $\logg$ with such high precision
based solely on spectra \citep{Meszaros-2013}. Although this relation is subject to systematic errors in \citet{Luck-2018}, we can robustly obtain $\logg$ that would allow the abundance measurements to be consistent with the results in \citet{Luck-2018}. The derived stellar parameters are found in detail in Appendix~\ref{AppendixA}.

Therefore, once $\TLDR$ gets estimated, we can use the relation (\ref{eq:tefflogglogp}) to estimate the surface gravity  of classical Cepheids.
Let us denote thus-estimated surface gravity as $\loggtrend$. 
For the precision of $\loggtrend$, we need to take into account the temperature error. A moderately large error of 100\,K in $\TLDR$ would lead roughly to the error of 0.12\,dex in $\loggtrend$ combined with the scatter of the $\Teff$--$\logg$--$\log P$ relation.
Using $\loggtrend$ is, therefore, more accurate and robust
than using $\loggLDR$.

\section{Line selection and calibration}
\label{sec:lines}

\subsection{Line Selection}
\label{sec:line_selection}
\begin{figure*}
    \centering \includegraphics[width=\textwidth]{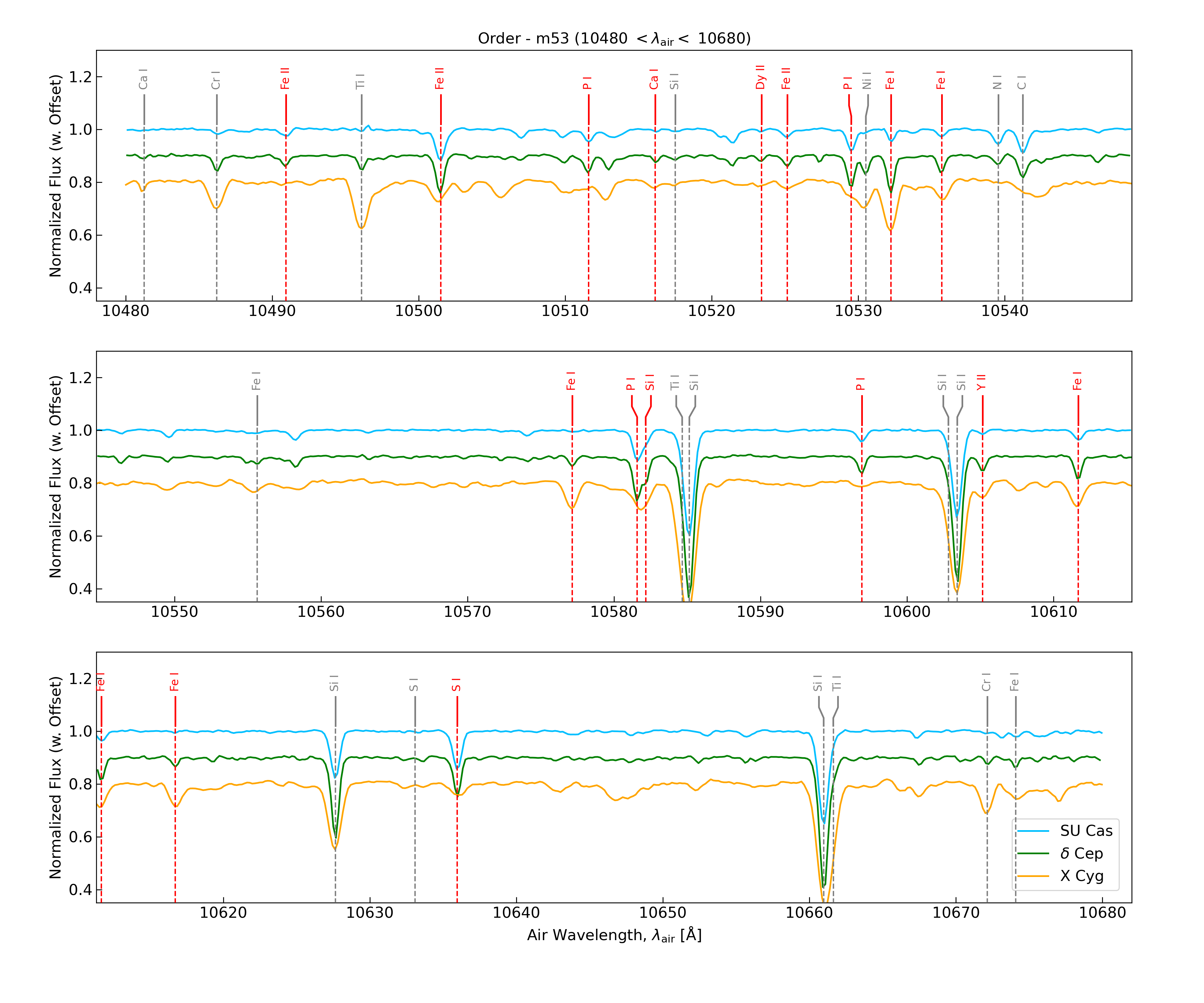}
    \caption{Species found in order 53 (Y band) for three calibrators, SU Cas, $\delta$ Cep, and X Cyg. Lines in gray correspond to those found in \citet{Fukue-2021}, while those in red correspond to those newly found in this study.}
    \label{fig:53-lines}
\end{figure*}

In this section, we discuss the line list to use for the abundance analysis. The first step is to select lines that are useful for measuring the abundances of different elements in the atmosphere of Cepheids. Concerning \ion{Fe}{i}, we adopt the list of lines selected  from VALD3 \citep{VALD3-15} and MB99 \citep{Melendez-1999} that were compiled by \citet{Kondo-2019}.  We also include other elements, among these lines we have \ion{Si}{i}, \ion{S}{i}, \ion{Ca}{i}, \ion{Ca}{ii}, \ion{Fe}{ii}, \ion{Zn}{i}, \ion{Y}{ii}, and \ion{Dy}{ii} from VALD and MB99 line lists. Most of the lines were confirmed and validated by examining the time-series high-resolution spectra we had available for this study. Figure~\ref{fig:53-lines} illustrates the absorption lines of different elements in our sample, the extended atlas of identified lines can be found in the supplementary online material.

For the species other than , we searched for candidate lines using synthetic spectra with MOOG \citep{Sneden-2012} considering a set of typical stellar parameters for Cepheids ($\Teff =5900$ K, $\logg = 2.0$, $\FeH=0.0$). We employed the stellar atmosphere models from MARCS \citep{Gustafsson-2008} ($\XH{M} \in [-2.5,0.5]$) spherical geometry for 5 $M_\odot$. The solar abundances in MARCS models are those of \citet{Grevesse-2007}, we converted them to the \citet{Asplund-2009} scale. Potential candidates were then filtered according to the following criteria:

\begin{itemize}
    \item They have to be deep enough ($d>0.03)$
    \item Blending ratio (the fraction of the EW of contaminating lines, $\beta _1$, defined in \citet{Kondo-2019}) less than 0.3.
    \item Having a wide ($\Teff$, $\logg$) range, $\Delta \Teff = 1000 K$, $\Delta \logg = 1.0$.
\end{itemize}

 The selection of lines resulted in 75 lines from VALD and 64 lines from MB99, and for the species other than \ion{Fe}{i}. Including \ion{Fe}{i}, the line list contains 105 lines from VALD and 90 lines from MB99. 

We consider the line information obtained from VALD and MB99 separately. If the same line appears in both lists, we calibrate its $\log gf$ and measure the abundance using the line twice. In the analyses with each line list, we use each list for all metallic lines the atomic lines, including other species.

\subsection{Line-by-line abundance}
\label{subsec:line-by-line}

The abundance inferred from an individual line is estimated by iteratively searching for the set of free parameters, including the abundance of interest, line broadening, wavelength shift, and continuum normalization factor that results from the synthetic spectrum reproducing the observed spectrum with the minimal residual. We employ the OCTOMAN code (Taniguchi et al. in prep.), which performs such optimization based on the MPFIT algorithm by \citet{Takeda-1995}. This algorithm has been utilized in various applications, including studies conducted with WINERED spectra \citep[]{Kondo-2019, Fukue-2021}. 

The utilization of the OCTOMAN tool for estimating abundances may encounter difficulties in certain scenarios. For instance, when blended lines are present but not accurately represented in the synthetic spectra, the fitting process may be compromised, resulting in unreliable estimates of the parameters including the abundance. In such cases, it is appropriate to reject the measurements. For each absorption line, we measured the abundance as a function of microturbulence. We considered a grid with 24 different $\xi$ values ranging from 1.4 to 6.0 km s$^{-1}$ with the step of 0.2 km s$^{-1}$. Each run of OCTOMAN was performed for a fixed $\xi \in [1.4-6.0]$
for each spectrum. 

\begin{figure}
    \centering
    \includegraphics[width=\columnwidth]{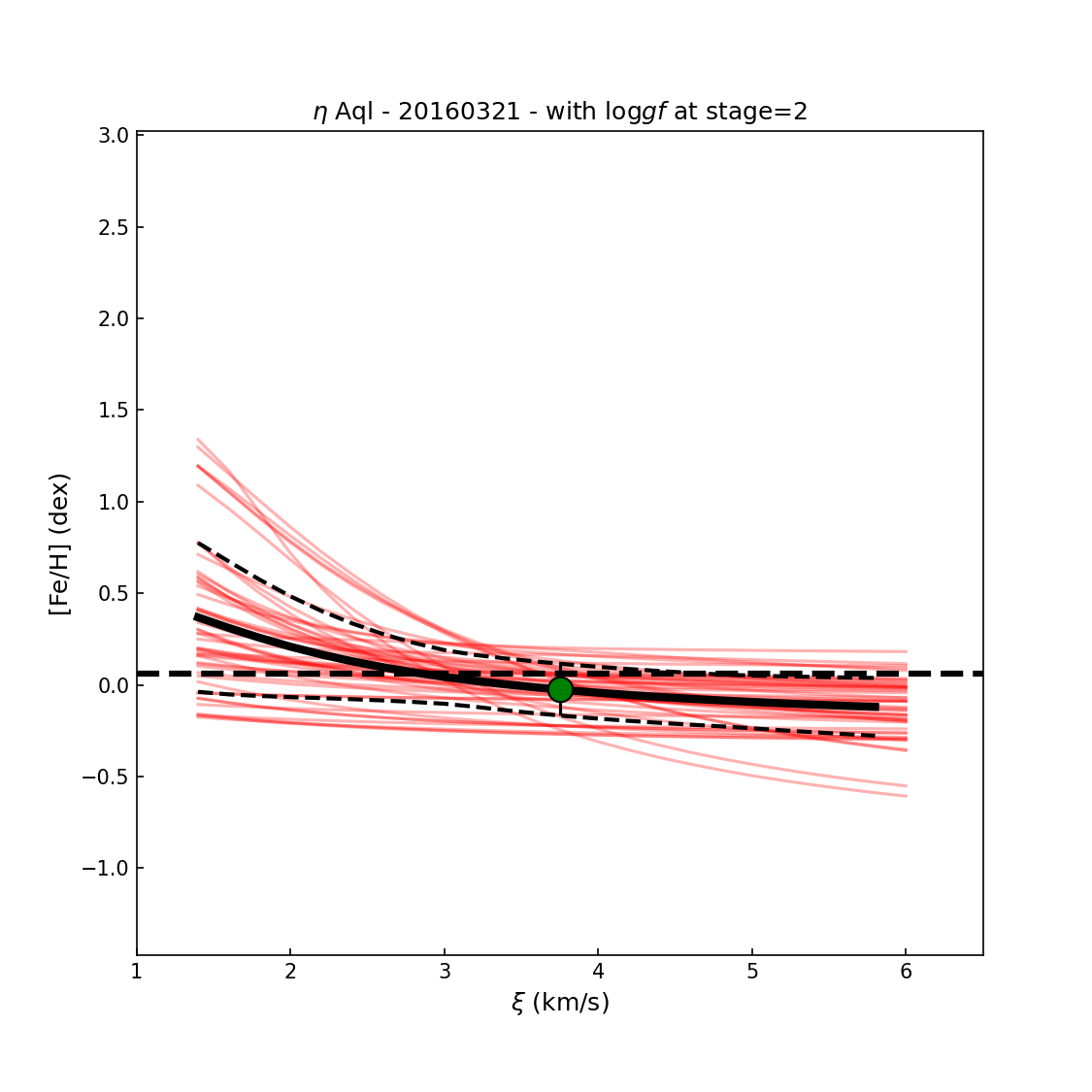}
    \caption{\ion{Fe}{i} abundances as a function of microturbulence. The case of $\eta$~Aql.}
    \label{fig:Eta_Aql_20160321}
\end{figure}

Due to the effect of line saturation, the abundance obtained in each OCTOMAN run depends on $\xi$. As illustrated in the example in Figure~\ref{fig:Eta_Aql_20160321}, different lines show
different trends of abundance against $\xi$. Deep lines show higher dependency on $\xi$
because they are more saturated, while weaker lines show little dependency. 
When many lines are strong and show the dependency on $\xi$,
the averaged abundance itself depends on $\xi$, i.e., the estimates of
abundance and $\xi$ show degeneracy. 
In contrast, if the majority of the lines are weak and depend little on $\xi$,
the final estimate of abundance is more-or-less independent of the estimated $\xi$.

In some cases, the OCTOMAN failed to give reasonable abundances for a part of the $\xi$ grid points. The rejection criteria for the OCTOMAN outputs consider the parameters, niter, fwhmv,  rv, and cnorm. The OCTOMAN measurements that passed the mentioned criteria are considered good when we have 19 or more grid points of $\xi$ for each line. Also, when we encounter $\xi - \XH{X}$ curves displaying a significant upturn by more than 0.05 dex, that means, if we find that the value of $\XH{X}$ at the largest grid point of $\xi$ is higher than the $\XH{X}$ at the smallest $\xi$ by 0.05 dex or more, that curve will be rejected. Thus, we accept curves having a sufficient number of good measurements showing a flat or decreasing trend over $\xi$. Otherwise, we reject the measurements of a given line
entirely if we get less than 20 measurements over the $\xi$ grid. 

The accepted curves for \ion{Fe}{i} lines are used when we determine $\xi$ in Section ~\ref{sec:method-xi}. 
Also, for other species, examining such trends is useful to see how
the microturbulence affects the measured abundance.
Moreover, we use a plot like Figure~\ref{fig:Eta_Aql_20160321} to identify
the absorption lines whose measurements for a particular spectrum need to be rejected.
We calculate the mean and standard deviation, $\sigma$, at each $\xi$ grid point. The solid black curve in Figure~\ref{fig:Eta_Aql_20160321} indicates 
the curve of mean, while the dashed curves  
indicate the upper and lower limits (the mean $\pm 2\,\sigma$).
Because of the different dependency on $\xi$, some curves of good measurements
(especially the lines showing strong dependency on $\xi$)
may well get outside the range between the upper and lower limits at some $\xi$. However, we reject the absorption lines whose curve in the $\xi$--[X/H] diagram
is outside the limits at all the $\xi$ grid points. 
We make this rejection of the outlying curves once
and re-calculate the mean and standard deviation at each $\xi$.

\subsection{Calibrating the oscillator strengths}
\label{sec:cal_loggf_fe1}
We calibrate the $\loggf$ values of absorption lines present
in Cepheids using our spectra of the calibrators. We assume that the stellar parameters including 
the microturbulence ($\xi$) are known for each phase of spectroscopic observation.
For the calibrators, as mentioned above, we obtained the stellar parameters ($\Teff, \logg, \xi$)
at each phase of our observations with the WINERED
by interpolating the curve of each parameter given by
\citet{Luck-2018}. 
For each combination of absorption line and spectrum,
we run the OCTOMAN to get the abundance.
The deviation $\Delta \loggf$ of this abundance from
the known abundance \citep{Luck-2018} includes, together with other errors,
the offset in $\loggf$.

\begin{figure}[ht]
  \centering
  \begin{minipage}[b]{0.45\textwidth}
    \centering
    \includegraphics[width=\textwidth]{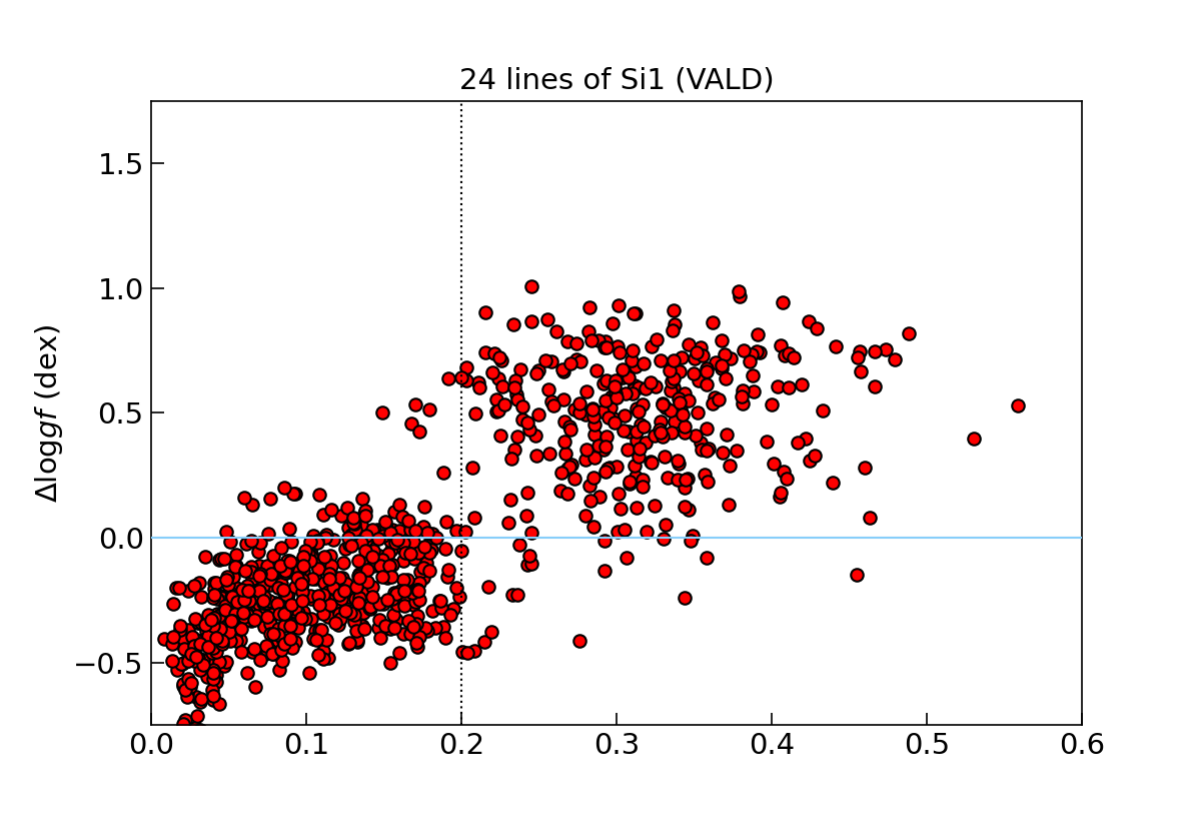}
    \caption{$\Delta\loggf_i$ for \ion{Si}{i} lines from VALD.}
    \label{fig:depth_delta_Si1_VALD}
  \end{minipage}
  \hfill
  \begin{minipage}[b]{0.45\textwidth}
    \centering
    \includegraphics[width=\textwidth]{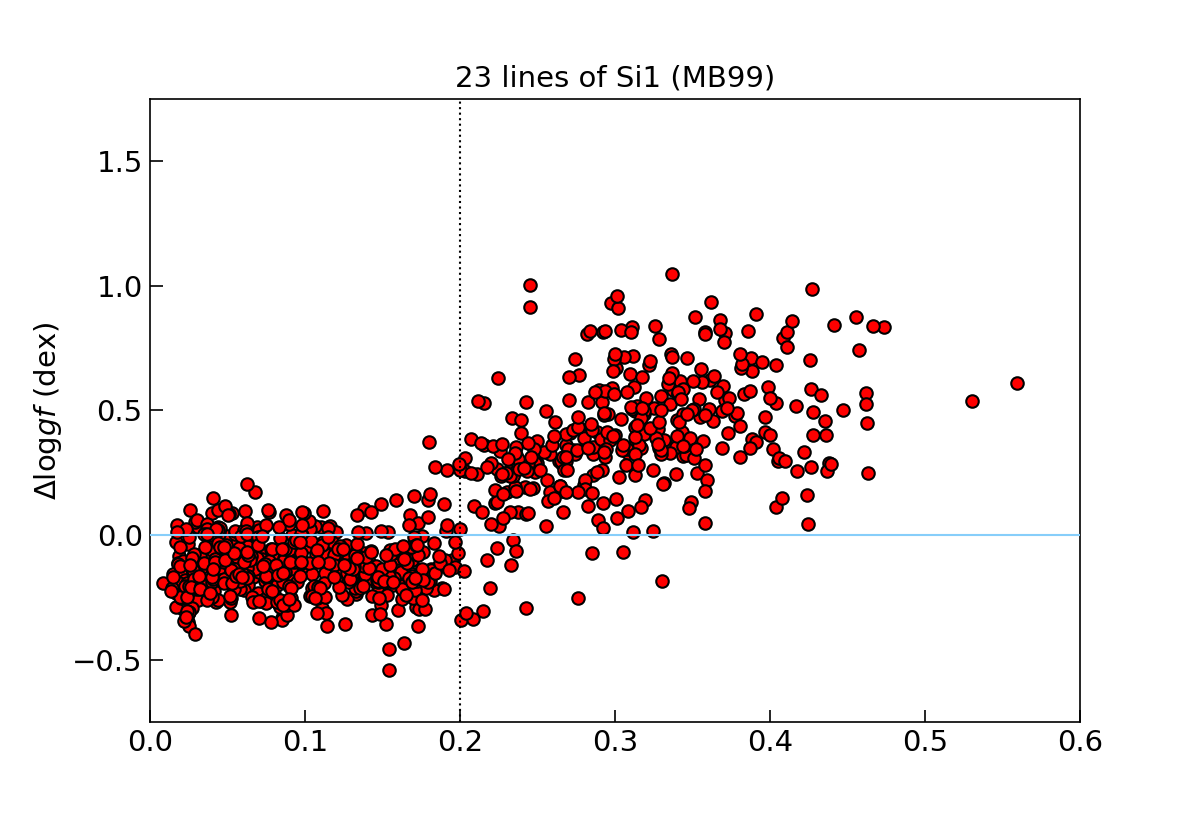}
    \caption{$\Delta\loggf_i$ for \ion{Si}{i} lines from MB99.}
    \label{fig:depth_delta_Si1_MB99}
  \end{minipage}
\end{figure}

We find that the $\Delta \loggf$ values obtained for deep lines
tend to show large systematic offsets.
This trend is clearer in the elements with both weak lines and
very strong lines like \ion{Si}{i}, more than \ion{Fe}{i}. Figures~\ref{fig:depth_delta_Si1_VALD}--\ref{fig:depth_delta_Si1_MB99} show
all the available $\Delta\loggf_i$ values for the case of neutral Si. 
The $\Delta\loggf_i$ values tend to be positive at depths larger than 0.2 - 0.25.
This may be partly attributed to a systematic trend buried 
in the original $\loggf$ values in VALD and MB99. 
However, such a systematic trend of $\Delta\loggf$ exceeding 0.5\,dex
for all strong lines is unexpected. It is instead understood as 
the limitation of using fixed microturbulence.
As we discuss in more detail in Section~\ref{sec:conclussion},
the microturbulence is expected to be significantly larger 
in the upper layers of the stellar atmosphere. 
Trying to calibrate $\loggf$ with underestimated $\xi$ would
lead to overestimates as we see here. We thus use
the measurements for lines shallower than 0.2 in depth
(the separation of the absorption core from the continuum level
in each observed spectrum).
In addition, among $\Delta \loggf$ of the not-too-strong lines,
we reject outliers with the $2\,\sigma$ clipping according to the standard deviation. 

Without including the measurements of lines whose depths were more than 0.2, we demand having more than 20 good measurements for each line in order to proceed with the calculations of new $\loggf$. The accepted measurements made a toll of 97 lines for VALD and 86 lines for MB99. Subsequently, OCTOMAN was run for each of the accepted lines in order to obtain the abundance of each of the elements, let us denote it as $\XH{X}$, as a function of $\xi$. 

For the accepted $\xi - \XH{X}$ curves, we estimate the $\XH{X}$ at the $\xi$ given (expected) from the \citet{Luck-2018} dataset (Section~\ref{sec:stellar_param}), and thus, we can obtain the necessary shift to match the literature $\XH{X}_{lit}$ values. We denominate such an offset given by each spectrum ($\Delta\loggf _i$), and calculate the mean of it from all the spectra that give accepted measurements for the particular line. In order to calculate the mean we need 20 or more $\Delta\loggf _i$ values. The final error of the calibrated $\loggf$ of this offset is obtained by calculating the standard deviation of the $\Delta\loggf _i$. This calibration was effective enough for 42 VALD lines and 37 MB99 lines for \ion{Fe}{I}, and for 48 VALD lines and 37 MB99 lines for the species different that \ion{Fe}{I}. The results of this procedure are given in Table~\ref{tab:calib_Fe1_loggf} and Table~\ref{tab:calib_species_loggf} given in Appendix~\ref{AppendixB}.

\begin{figure*}[H]
    \centering
    \includegraphics[clip,width=\hsize]{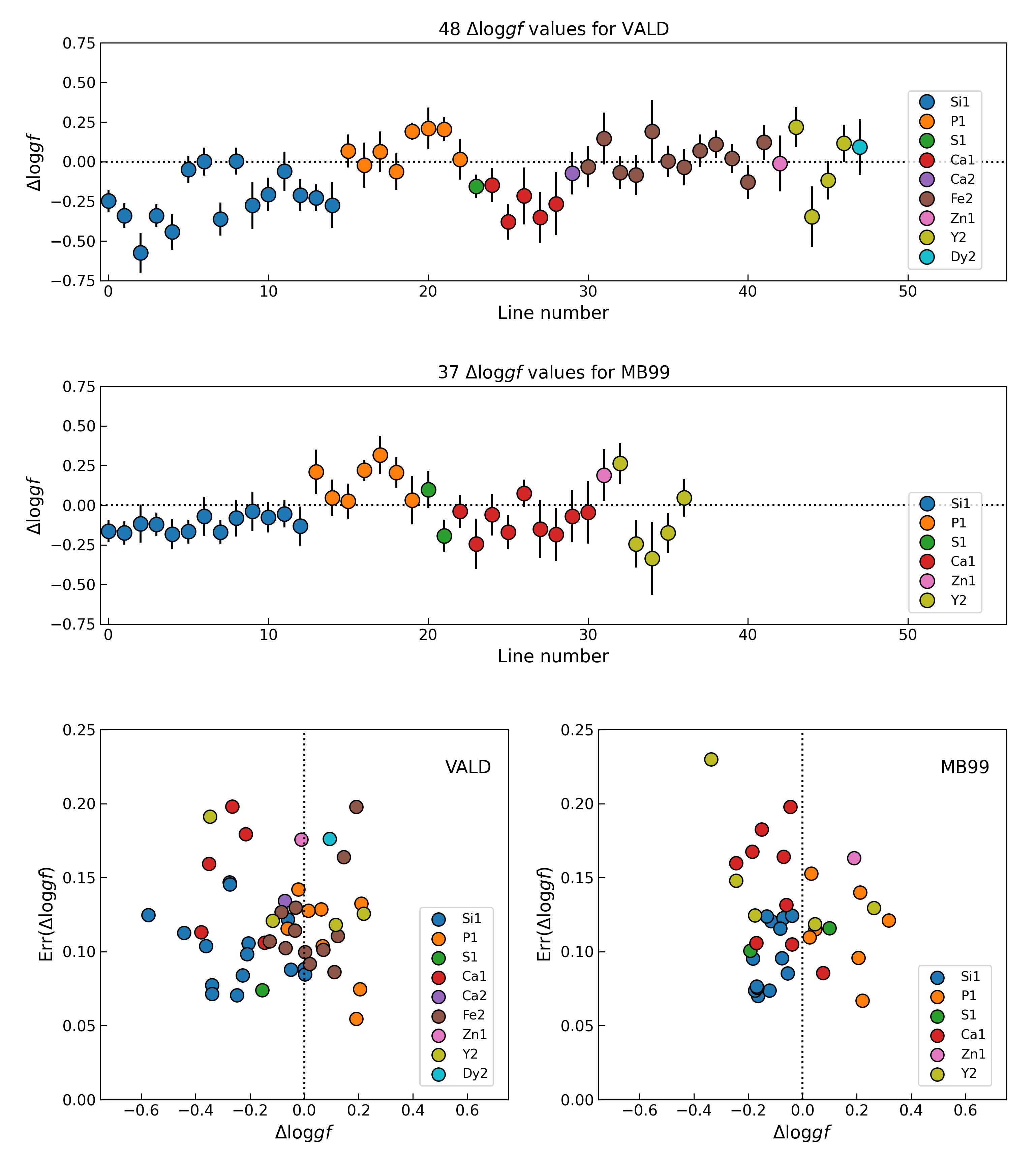}
    \caption[The offsets between the original and new $\loggf$ values in our calibration]{The offsets between the original and new $\loggf$ values in our calibration for the lines taken from VALD (upper) and MB99 (middle).}
    \label{fig:line_deltas_species}
\end{figure*}

\section{Abundance Analysis}
\label{sec:ab_analysis}
\subsection{Determining microturbulence}
\label{sec:method-xi}
In this section, we demonstrate the abundance analysis using the calibrated $\loggf$ but without using the stellar parameters estimated with the measurements by \citet{Luck-2018}. Our analysis starts by estimating $\TLDR$ and $\loggtrend$ as described in Section ~\ref{sec:stellar_param} and moves on to the determination of $\xi$ before the abundance measurements.
The basic idea for determining the microturbulence $\xi$ is to find the condition
that makes the abundances from lines of various strengths
(non- or less-saturated lines to saturated lines) consistent.
\citet{Mucciarelli-2011} discussed some methods of determining $\xi$. We measure the abundance from each line
as a function of $\xi$. Combining the results for lines with different strengths, we search for $\xi$ with which the abundances show no dependency on line strength. We use the $X$ index:
\begin{equation}
     X = \loggf -  EP \cdot \frac{5040}{0.86\,\Teff} \label{eq:Xindex}
\end{equation}

as the indicator of line strength \citep{Magain-1984,Gratton-2006, Kondo-2019}.
The equivalent width might be a good indicator, but 
measuring the equivalent width in the observed spectra is not a trivial task as other factors, such as blending, can easily disturb the results.
We measure the line depth and use it for rejecting lines deeper than 0.2 in depth, but prefer to use the $X$ indicator here because it is a theoretical value unaffected by
observational errors. 
Then, at each $\xi$, we fit the linear relation,\\

\begin{equation}
    \FeH = aX + b, \label{eq:xitrend}
\end{equation}
\\
to the abundances ($\FeH$) from individual lines as a function of the $X$ index.
The slope $a$ changes, usually in a monotonic way, with $\xi$ and we select the $\xi$, one of the $\xi$ grid points
with the $0.2\,\kms$ step, that gives the slope closest to zero.
\citet{Fukue-2021} performed a simulation to find that determining $\xi$ requires
at least 20 absorption lines. We use only \ion{Fe}{i} lines to estimate $\xi$.

\subsection{Abundances with a given microturbulence} 
\label{sec:mean_X}

Once the microturbulence is obtained, it is relatively simple to derive the abundances making use of an established list of lines (with calibrated $\loggf$). 
For each species, we calculate the weighted mean of $\XH{X}_{i}$ 
from $N$ individual lines that were not rejected, using OCTOMAN to calculate $\XH{X}_{i}$ for individual lines and took the weighted mean,
\begin{equation}
\XH{X} = \left. \sum_{i=1}^N \left(w_i \XH{X}_i \right) \right/ \sum_{i=1}^N w_i ,
\label{eq:mean_X}
\end{equation}
where the weights are given by $w_i=1/e_i^2$ and $e_i$ is 
the error in the calibrated $\loggf$. 
We consider the weighted standard deviation,
\begin{equation}
e_\mathrm{X,1} = \sqrt{\left. \sum_{i=1}^N w_i \left(\XH{X}_i - \XH{X}\right)^2 \right/ \sum_{i=1}^N w_i},
\label{eq:std1}
\end{equation}
as the error of $\XH{X}$.
Using the standard deviation rather than the standard error (i.e., the standard deviation divided by $\sqrt{N-1}$) is commonly done 
and recommended by \citet{Jofre-2019}. 

However, Equation~(\ref{eq:std1}) would underestimate
the uncertainty of the averaged abundance when 
$\XH{X}_i$ with a small number of lines get
closer, than expected from the statistical errors
to each other by coincidence.
We consider another indicator of uncertainty,
\begin{equation}
e_\mathrm{X,2} = \sqrt{1 \left/ \sum_{i=1}^N w_i\right.} ~,
\label{eq:std2}
\end{equation}
which is given by the error propagation. 
Then, we take the larger of $e_\mathrm{X,1}$ and $e_\mathrm{X,2}$
as the error $e_\mathrm{X}$ of the derived $\XH{X}$. The results of this analysis can be found in Appendix~\ref{AppendixC}.

When we have $\XH{X}$ estimated with multiple spectra of each Cepheid,
we would like to take the mean of phase-by-phase measurements. 
We calculate the weighted mean using the formula like
Equation~(\ref{eq:mean_X}) but with the weights that are determined
with the $e_X$ of individual phases. 
The errors are also determined in the same way
by taking the larger of the weighted standard deviation (Equation~\ref{eq:std1}) and the propagated error (Equation~\ref{eq:std2})

\subsection{Comparison with the abundances in Luck+18}

Using the 46 VALD and 35 MB99 lines with $\loggf$ calibrated,
we determined the abundances, $\XH{X}$, of 9 species.
The results are given in Appendix~(number).
Figures~\ref{fig:result-Si1} plot $\XH{X}$ and  the deviations from the literature, \citet{Luck-2018},
for the calibrators and the validators. 
For a few species, like \ion{Si}{i} (Figure~\ref{fig:result-Si1}),
we find very good agreements between our measurements and the literature values ($\sigma \sim 0.07$ dex). 
Some species, however, show large scatters ($\sigma > 0.15$ dex).

\begin{figure*}[htbp]
    \centering
    \includegraphics[width=\textwidth]{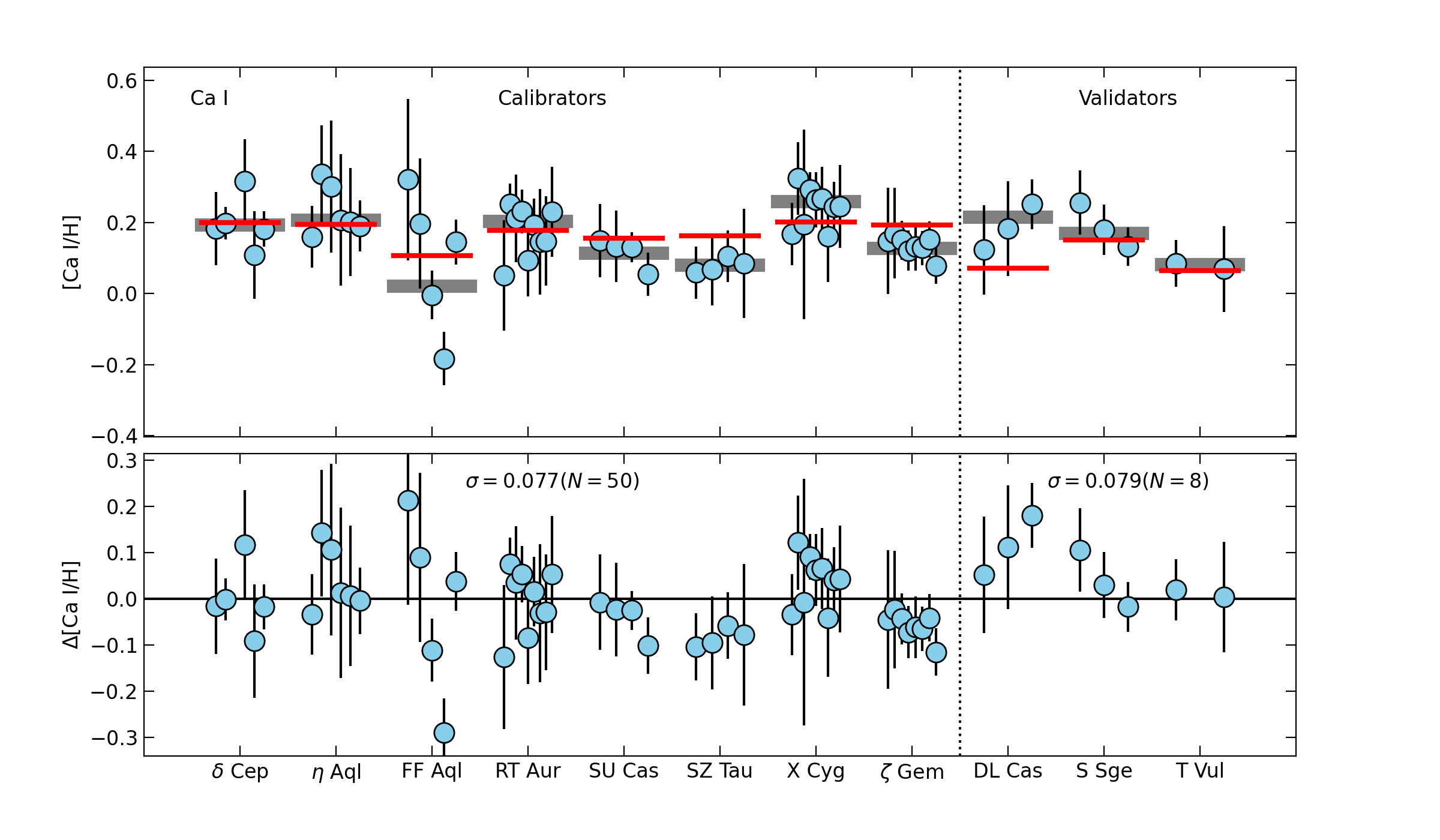}
    \caption{The derived abundances (upper) and the deviations from the literature values (lower) for \ion{Ca}{i}.}
    \label{fig:result-Ca1}
\end{figure*}

Table~\ref{tab:species_scatters} summarizes how well the abundances we derived agree with
the literature values.
The mean offsets from the abundance scale of \citet{Luck-2018}
are not significant for both calibrators and validators. This agreement is expected, at least for the calibrators, because we calibrated $\loggf$ based on the abundances of \citet{Luck-2018}.
Yet, the standard deviation is large for some species, in particular, \ion{Ca}{ii} and \ion{Zn}{i}. 
The large SDs may be partly attributed to errors in \citet{Luck-2018}. No significant offsets are found (compared to SD). Looking at Figure~\ref{fig:result-Ca1} for \ion{Ca}{i} and Figure~\ref{fig:result-Si1} for \ion{Si}{i}, 
the abundances of some objects from \citet{Luck-2018} seem to show unexpected offsets
from other objects; see, e.g., X~Cyg for \ion{Zn}{i}. 

\begin{figure*}[htbp]
    \centering
    \includegraphics[width=\textwidth]{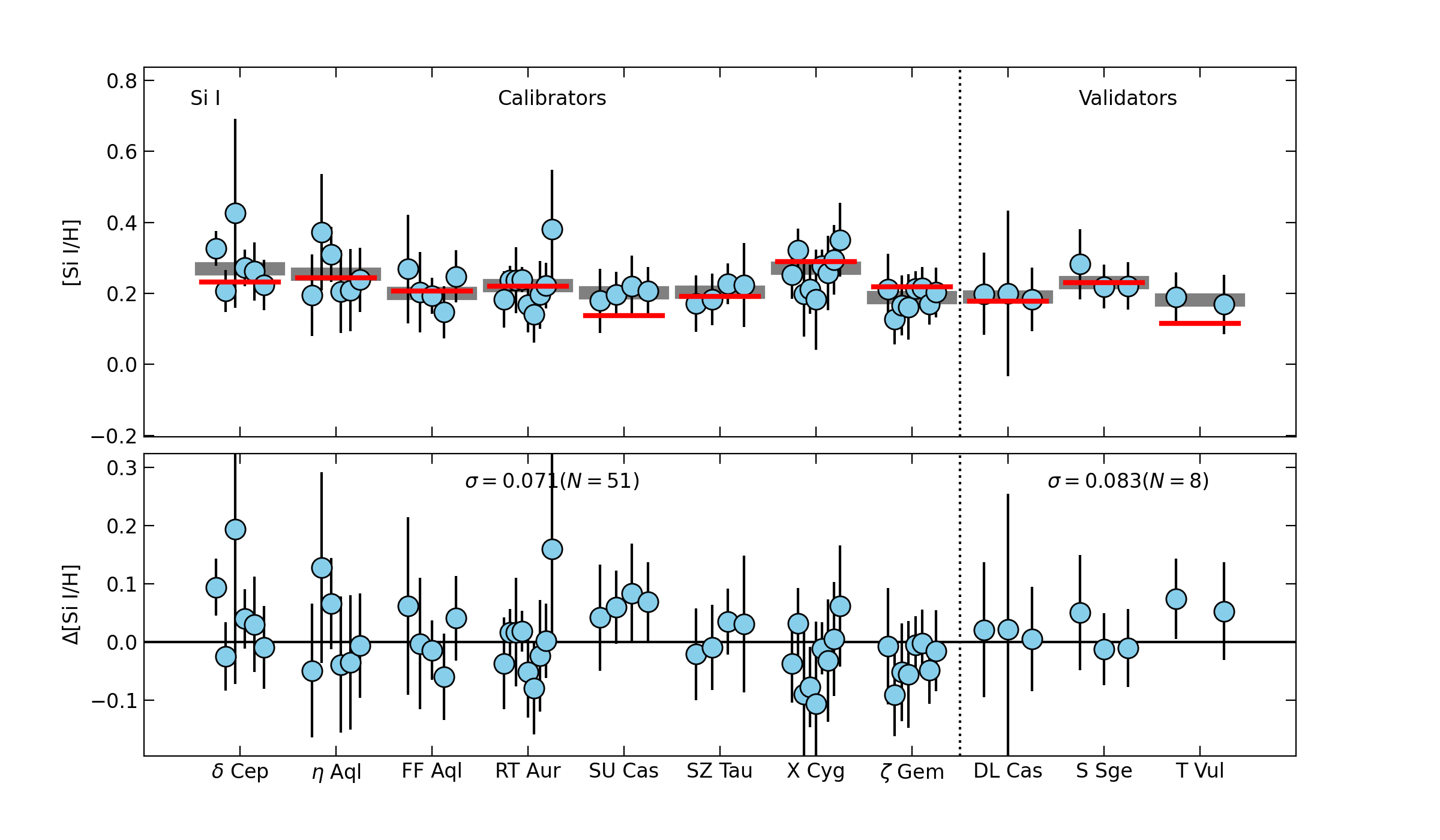}
    \caption{The derived abundances (upper) and the deviations from the literature values (lower) for \ion{Si}{i}.}
    \label{fig:result-Si1}
\end{figure*}

\begin{table}
\caption{Deviations of abundances from \citet{Luck-2018} for species other than Fe I \label{tab:species_scatters}}
\begin{center}
\begin{tabular}{lrrrrrr}
\hline\hline
 Species & \multicolumn{3}{c}{Calibrators} & \multicolumn{3}{c}{Validators} \\
 & \multicolumn{1}{c}{Mean} & \multicolumn{1}{c}{SD} & \multicolumn{1}{c}{$N$} & \multicolumn{1}{c}{Mean} & \multicolumn{1}{c}{SD} & \multicolumn{1}{c}{$N$} \\
 & \multicolumn{1}{c}{(dex)} & \multicolumn{1}{c}{(dex)} & & \multicolumn{1}{c}{(dex)} & \multicolumn{1}{c}{(dex)} & \\
\hline\hline
$\mathrm{Si\,I}$  & $0.003$ & 0.071 & 51 & $0.003$ & 0.083 & 8 \\
$\mathrm{P\,I}$   & $-0.003$ & 0.064 & 51 & $-0.003$ & 0.065 & 8 \\
$\mathrm{S\,I}$   & $-0.029$ & 0.100 & 51 & $-0.029$ & 0.108 & 8 \\
$\mathrm{Ca\,I}$   & $-0.022$ & 0.077 & 50 & $-0.022$ & 0.079 & 8 \\
$\mathrm{Ca\,II}$  & $-0.051$ & 0.268 & 32 & $-0.051$ & 0.209 & 4 \\
$\mathrm{Fe\,II}$  & $-0.006$ & 0.072 & 51 & $-0.006$ & 0.090 & 8 \\
$\mathrm{Zn\,I}$  & $-0.080$ & 0.179 & 37 & $-0.080$ & 0.129 & 4 \\
$\mathrm{Y\,II}$  & $-0.003$ & 0.114 & 48 & $-0.003$ & 0.090 & 8 \\
$\mathrm{Dy\,II}$  & $-0.078$ & 0.176 & 14 & $-0.078$ & 0.176 & 4 \\
\hline
\end{tabular}
\end{center}
\end{table}

\section{Concluding remarks} 
\label{sec:conclussion}
We have successfully established a list of elements in neutral and ionized state of nine species (including Fe) that are useful for measuring the abundances of classical Cepheids. 
We calibrated the $\loggf$ of these lines by comparing our measurements with the abundances of eight calibrators in \citet{Luck-2018},  and thus-calibrated $\loggf$ gives the abundances of the validators consistent with those in \citet{Luck-2018}. The precision for a few species is as high as the abundance analysis with optical spectra.  A few points are worth further consideration.

(1) The method described in this work for obtaining effective temperatures using the LDR relations leads to high precision ($\leq$ 100 K) except for some cases with high $\Teff$ ($\geq$ 6000 K) and/or low S/N ($\leq$ 150). This gives a rough guide to future observers; $\textit{YJ}$ - band spectra with S/N $\sim$ 100 or higher would allow precise abundance measurements, while warmer Cepheids would require higher S/N.

(2) The impact of microturbulence on the strength of each line is known to be affected by the line strength. In the determination of $\loggf$ of a given species, a question arises as to whether it is appropriate to use $\xi$ obtained solely with \ion{Fe}{i} lines. We have observed so that the utilization of stronger lines, such as those of Silicon or Strontium, may also impact the resulting $\xi$, providing evidence that $\xi$ changes along the atmosphere of Cepheids at different $\tau$. This phenomenon was initially reported by \citet{Takeda-1997} in an HB star, and later on in RR Lyrae by \citet{Takeda-2006, Kolenberg-2010, Fossati-2014}. This suggests that $\xi$ increases with height in the atmosphere of RR Lyrae stars, as reported in \citet{Fossati-2014}. Additionally, it is recommended that a specific $\xi$ be applied to lines with specific strength to avoid erroneous abundance estimates. 
Considering the depth-dependent $\xi$ could make it possible to
include such strong lines, but we have no set of atmosphere models
with depth-dependent $\xi$ for Cepheids in which the turbulent velocity field may well be different from static stars. Therefore, in this paper, we rejected lines stronger than 0.2 and focused on weaker lines and we suggest to avoid such strong lines in the future abundance analyses. 

(3) Phosphorus, an odd-Z ($Z=15$) element, whose isotope belongs to the same group as Nitrogen is believed to have been produced during the C and Ne burning. It is one of the most fundamental life-bearing elements dominated by macromolecules composed mainly of Carbon that are present in every single organism formed by cells. Each contains membranes and inner organelles whose primary component is a molecule denominated Phospholipid (a.k.a.\ phosphatides) which is a class of lipid whose molecule is formed by a hydrophilic head where the phosphate group is located as well as two hydrophobic tails. Moreover, from Biology, we know that when DNA is synthesized, a molecule called adenosine triphosphate (ATP) brings energy to the whole DNA structure, by keeping it together, as a superglue.
Phosphorus, chemically-wise, has been surprisingly skipped for years. An example of this is that prior 2011, with the study of \citet{Caffau-2007}, P was barely analyzed in stars. A possible cause for this sort of dodging was raised back in the early 1930's by \citet{Struve-1930} who found that no neutral P was available in spectra of stars whose spectral type
ranged from F-K. Though, it was acknowledged that Phosphorus ions might be visible
in other kinds of stars or other regimes \citep{Maas-2020, Naghma-2018}, such as \ion{P}{ii}, \ion{P}{ii}, \ion{P}{iii}, \ion{P}{iv} observable in UV
spectra. 
A handful of studies have been done on stars after the study of \citet{Caffau-2007} which first suggested using the high-excitation \ion{P}{i} lines in multiplet at 1051-1068 nm. Different authors have focused their attention on various other targets of interest: such as \citet{Hubrig-2009} obtained \ion{P}{i} abundance of HB stars in globular clusters. \citet{Melendez-2009-10} investigated the \ion{P}{i} abundances of a handful of solar twins. \citet{Sbordone-2009} suggested that Sulphur could have been produced by \textit{proton capture} on P, leading to a debate on how possible it is to create the amount of S found there as a product of P when P was never measured on stars in that globular cluster. The work on stars in our galaxy was first reported with the P abundance on 22 MS (main sequence) stars \citet{Caffau-2005}, stars in the galactic disk \citet{Caffau-2011}, and dwarf stars \citep{Caffau-2016}. Discussions on biological and geological implications can be found, e.g., in \citet{Hinkel-2020}. 
We detected more lines of \ion{P}{i} present in our NIR spectra observed with WINERED whose abundances were also measured. None of the aforementioned studies have included Cepheids. Our study on the P abundance is probably the first to be performed on stars as hot and variable as Cepheids. We confirmed ten \ion{P}{i} lines in our spectra of which 9 had their $\loggf$ values calibrated. 

(4) From the works of \citet{Luck-2018} on hundreds of Cepheids, which itself included our sample of Calibrators and Validators, the abundances of a handful of elements were provided. Our estimates of chemical abundances are in agreement with the abundance gradient found by \citet{Luck-2018} and prove that we can use spectra in the $YJ$ bands to take the study of the Galactic Chemistry to the next level by including the interstellar-hidden Cepheids that can be observed in the infrared only.

\section*{Acknowledgments}
We are grateful to the staff of the Koyama Astronomical
Observatory for their support during our observation. This study
is supported by JSPS KAKENHI No.18H01248 and No.19KK0080. The WINERED spectrograph was developed by the University of
Tokyo and the Laboratory of Infrared High-resolution Spectroscopy
(LiH), Kyoto Sangyo University under the financial support of
KAKENHI (numbers 16684001, 20340042, and 21840052) and the
MEXT Supported Program for the Strategic Research Foundation at
Private Universities (numbers S0801061 and S1411028).
SE acknowledges the financial support of Millenium Nucleus ERIS NCN2021$\_$017, and the ANID Millennium Institute of Astrophysics MAS (ICN)12.009. DT is financially supported by JSPS Research Fellowship for Young Scientists and accompanying Grant-in-Aid for JSPS Fellows,  DT (No. 21J11555). DT also acknowledges
financial support from Masason Foundation.

\vspace{5mm}




\clearpage
\appendix

\section{Stellar Parameters}
\label{AppendixA}
Here, we provide the derived stellar parameters ($\Teff$ and $\logg$) for each of the targets at each of their obtained phases.

\begin{table}[H]
\caption{Stellar Parameters derived for the calibrators sample.}
\centering
\begin{tabular}{lccccccccc}
\toprule
ID &  Phase &   $\TLDR$ &  $e_{\TLDR}$ &  $N_{\TLDR}$ &  $\loggLDR$ &  $e_\loggLDR$ &  $N_{\loggLDR}$ &  $\loggtrend$ &  $e_{\loggtrend}$\\
\toprule
$\delta$ Cep &  0.119 &  6158.858 &  109.492 &            9 &       1.995 &           0.129 &              10 &       2.079 &           0.119 \\
             &  0.178 &  5974.720 &   95.670 &           10 &       2.177 &           0.163 &               8 &       1.993 &           0.117 \\
             &  0.556 &  5564.740 &   52.993 &           12 &       1.974 &           0.118 &               9 &       1.793 &           0.111 \\
             &  0.672 &  5460.668 &   78.134 &           11 &       1.611 &           0.140 &              10 &       1.740 &           0.115 \\
             &  0.711 &  5530.576 &   41.254 &           11 &       1.812 &           0.104 &               9 &       1.776 &           0.110 \\
             &  0.897 &  6305.770 &  292.722 &            3 &       2.310 &           0.194 &               6 &       2.145 &           0.170 \\
\hline
$\eta$ Aql &  0.174 &  5851.384 &   86.724 &           11 &       1.755 &           0.145 &               8 &       1.836 &           0.116 \\
           &  0.216 &  5846.917 &   75.231 &           10 &       1.757 &           0.106 &               9 &       1.834 &           0.114 \\
             &  0.395 &  5674.762 &   90.609 &           11 &       2.070 &           0.161 &               9 &       1.750 &           0.117 \\
            &  0.700 &  5304.119 &   41.614 &           11 &       1.475 &           0.107 &               9 &       1.560 &           0.110 \\
            &  0.743 &  5357.368 &  145.628 &            8 &       1.439 &           0.235 &               6 &       1.588 &           0.132 \\
            &  0.887 &  5886.794 &  134.170 &            9 &       1.774 &           0.147 &               8 &       1.853 &           0.126 \\
\hline
FF Aql &  0.105 &  6257.612 &   66.101 &            9 &       2.225 &           0.102 &              10 &       2.185 &           0.112 \\
        &  0.594 &  5918.142 &  148.542 &            6 &       2.067 &           0.172 &               7 &       2.028 &           0.129 \\
        &  0.627 &  6156.364 &   62.646 &           12 &       2.143 &           0.105 &              10 &       2.139 &           0.112 \\
        &  0.737 &  6299.516 &   54.178 &           10 &       2.294 &           0.097 &              10 &       2.204 &           0.111 \\
        &  0.823 &  6155.288 &  154.563 &            9 &       1.774 &           0.112 &              10 &       2.138 &           0.129 \\
\hline
 RT Aur &  0.149 &  6303.853 &  168.488 &            6 &       2.207 &           0.151 &               8 &       2.267 &           0.132 \\
        &  0.155 &  6382.660 &   81.993 &            9 &       2.385 &           0.109 &              10 &       2.302 &           0.114 \\
        &  0.252 &  6069.138 &   47.748 &           12 &       2.136 &           0.098 &              10 &       2.160 &           0.110 \\
       &  0.341 &  5851.989 &   44.895 &           12 &       1.940 &           0.097 &              10 &       2.057 &           0.110 \\
       &  0.422 &  5711.910 &   50.499 &           11 &       1.917 &           0.102 &              10 &       1.989 &           0.111 \\
       &  0.492 &  5680.734 &   45.042 &           12 &       1.760 &           0.100 &              10 &       1.974 &           0.110 \\
       &  0.620 &  5667.429 &   50.531 &            8 &       1.927 &           0.123 &               7 &       1.967 &           0.111 \\
        &  0.776 &  5737.308 &   81.643 &           12 &       1.913 &           0.159 &               7 &       2.001 &           0.115 \\
      &  0.920 &  6206.251 &   67.663 &            9 &       2.166 &           0.103 &              10 &       2.223 &           0.112 \\
\hline
SU Cas &  0.108 &  6341.471 &   98.568 &            8 &       2.193 &           0.114 &              10 &       2.502 &           0.117 \\
       &  0.150 &  6420.494 &  123.257 &            7 &       2.305 &           0.121 &              10 &       2.537 &           0.121 \\
      &  0.546 &  6103.989 &   75.763 &           11 &       2.261 &           0.112 &              10 &       2.394 &           0.114 \\
\hline
 SZ Tau &  0.502 &  5813.069 &   82.707 &           12 &       1.953 &           0.143 &               9 &       2.095 &           0.115 \\
        &  0.639 &  5799.423 &   56.040 &           12 &       1.860 &           0.112 &               9 &       2.089 &           0.111 \\
        &  0.772 &  6114.620 &  128.669 &            7 &       2.334 &           0.142 &               9 &       2.238 &           0.123 \\
        &  0.870 &  6099.724 &   87.401 &            6 &       2.034 &           0.111 &               9 &       2.231 &           0.115 \\
\hline

  X Cyg &  0.030 &  6006.767 &   61.034 &           11 &       1.794 &           0.114 &               8 &       1.632 &           0.112 \\
          &  0.103 &  5628.700 &   48.255 &           10 &       1.404 &           0.098 &              10 &       1.449 &           0.111 \\
           &  0.353 &  5058.509 &   39.302 &           11 &       1.385 &           0.117 &              10 &       1.149 &           0.110 \\
           &  0.408 &  4912.173 &   61.398 &            9 &       0.992 &           0.268 &               8 &       1.066 &           0.114 \\
          &  0.582 &  4716.821 &   51.199 &           10 &       0.739 &           0.466 &               5 &       0.952 &           0.112 \\
          &  0.408 &  4912.173 &   61.398 &            9 &       0.992 &           0.268 &               8 &       1.066 &           0.114 \\
\hline     
\end{tabular}
\end{table}        
\begin{table}[H]
\centering
\begin{tabular}{lccccccccc}
\toprule
ID &  Phase &   $\TLDR$ &  $e_{\TLDR}$ &  $N_{\TLDR}$ &  $\loggLDR$ &  $e_\loggLDR$ &  $N_{\loggLDR}$ &  $\loggtrend$ &  $e_{\loggtrend}$\\
\toprule       
X Cyg     &  0.582 &  4716.821 &   51.199 &           10 &       0.739 &           0.466 &               5 &       0.952 &           0.112 \\
       &  0.606 &  4726.672 &   42.517 &           11 &       0.759 &           0.208 &               4 &       0.958 &           0.111 \\
           &  0.685 &  4852.158 &   51.921 &           10 &       1.191 &           0.159 &               5 &       1.031 &           0.112 \\
          &  0.760 &  5101.901 &   70.059 &            9 &       1.080 &           0.233 &               2 &       1.173 &           0.115 \\
          &  0.880 &  5383.506 &   39.783 &           12 &       1.467 &           0.106 &               9 &       1.324 &           0.110 \\
\hline
$\zeta$ Gem &  0.028 &  5803.001 &   41.689 &           10 &       1.696 &           0.094 &              10 &       1.696 &           0.110 \\
             &  0.133 &  5636.552 &   41.616 &           12 &       1.579 &           0.102 &               9 &       1.614 &           0.110 \\
           &  0.200 &  5488.495 &   42.429 &           12 &       1.483 &           0.099 &              10 &       1.540 &           0.110 \\
            &  0.534 &  5246.328 &   36.864 &           12 &       1.365 &           0.100 &               9 &       1.412 &           0.110 \\
            &  0.626 &  5509.257 &   44.747 &           10 &       1.740 &           0.113 &               8 &       1.550 &           0.110 \\
            &  0.645 &  5442.997 &   38.323 &           11 &       1.433 &           0.101 &               9 &       1.516 &           0.110 \\
             &  0.791 &  5706.932 &   54.991 &           12 &       1.671 &           0.110 &               9 &       1.649 &           0.111 \\
           &  0.923 &  5711.450 &   43.379 &           10 &       1.508 &           0.095 &              10 &       1.652 &           0.110 \\
\hline
\end{tabular}
\end{table}

\begin{table}[H]
\caption{Stellar Parameters derived for the validators sample.}
\centering
\begin{tabular}{lccccccccc}

\toprule
ID &  Phase &   $\TLDR$ &  $e_{\TLDR}$ &  $N_{\TLDR}$ &  $\loggLDR$ &  $e_\loggLDR$ &  $N_{\loggLDR}$ &  $\loggtrend$ &  $e_{\loggtrend}$\\
\toprule
  DL Cas &  0.219 &  5855.625 &   94.278 &            8 &       1.924 &           0.124 &               9 &       1.802 &           0.117 \\
           &  0.604 &  5310.473 &   83.805 &            8 &       1.550 &           0.158 &               7 &       1.527 &           0.117 \\
           &  0.743 &  5422.091 &  107.744 &            8 &       1.662 &           0.229 &               3 &       1.585 &           0.122 \\
\hline
    S Sge &  0.119 &  5996.566 &   71.749 &           12 &       1.968 &           0.115 &              10 &       1.853 &           0.113 \\
          &  0.520 &  5342.743 &   39.696 &           12 &       1.426 &           0.098 &              10 &       1.528 &           0.110 \\
          &  0.940 &  6097.361 &  107.231 &           10 &       1.823 &           0.134 &               9 &       1.900 &           0.119 \\
\hline
    T Vul &  0.138 &  6131.199 &  102.707 &            9 &       2.104 &           0.135 &               8 &       2.130 &           0.118 \\
          &  0.293 &  5798.716 &   52.616 &           12 &       1.798 &           0.104 &              10 &       1.973 &           0.111 \\

\hline
\end{tabular}
\end{table}

\clearpage

\section{Calibrated log gf for Fe I and species lines}
\label{AppendixB}
Here, we include the results of our method for calculating the oscillator strengths ($\loggf$ values) by following the procedure stated in Section~\ref{sec:cal_loggf_fe1}.

{\small
\begin{longtable}{crrrrr}
\caption{Calibration of $\log gf$ of \ion{Fe}{i} lines \label{tab:calib_Fe1_loggf}} \\
\hline\hline
 \multicolumn{1}{c}{$\lambda_\mathrm{air}$} & \multicolumn{1}{c}{EP} & \multicolumn{2}{l}{$\log gf_\mathrm{VALD}$ (dex)} & \multicolumn{2}{l}{$\log gf_\mathrm{MB99}$ (dex)} \\
 \multicolumn{1}{c}{(\AA)} & \multicolumn{1}{c}{(eV)} & \multicolumn{1}{c}{old} & \multicolumn{1}{c}{new (SD, $N$)} & \multicolumn{1}{c}{old} & \multicolumn{1}{c}{new (SD, $N$)} \\
\hline\hline
\endfirsthead
\caption[]{Calibration of $\log gf$ of \ion{Fe}{i} lines---continued.} \\
\hline
 \multicolumn{1}{c}{$\lambda_\mathrm{air}$} & \multicolumn{1}{c}{EP} & \multicolumn{2}{l}{$\log gf_\mathrm{VALD}$ (dex)} & \multicolumn{2}{l}{$\log gf_\mathrm{MB99}$ (dex)} \\
 \multicolumn{1}{c}{(\AA)} & \multicolumn{1}{c}{(eV)} & \multicolumn{1}{c}{old} & \multicolumn{1}{c}{new (SD, $N$)} & \multicolumn{1}{c}{old} & \multicolumn{1}{c}{new (SD, $N$)} \\
\hline
\endhead
\hline
\endfoot
9800.3075 & 5.0856 & $-0.453$ & $-0.715$ (0.139, 21) & \multicolumn{1}{c}{---} & \multicolumn{1}{c}{---} \\
9811.5041 & 5.0117 & $-1.362$ & $-1.487$ (0.098, 33) & \multicolumn{1}{c}{---} & \multicolumn{1}{c}{---} \\
9861.7337 & 5.0638 & $-0.142$ & $-0.623$ (0.084, 39) & \multicolumn{1}{c}{---} & \multicolumn{1}{c}{---} \\
9868.1857 & 5.0856 & $-0.979$ & $-0.885$ (0.075, 39) & \multicolumn{1}{c}{---} & \multicolumn{1}{c}{---} \\
9944.2065 & 5.0117 & $-1.338$ & $-1.476$ (0.126, 22) & \multicolumn{1}{c}{---} & \multicolumn{1}{c}{---} \\
9980.4629 & 5.0331 & $-1.379$ & $-1.546$ (0.093, 43) & \multicolumn{1}{c}{---} & \multicolumn{1}{c}{---} \\
10041.472 & 5.0117 & $-1.772$ & $-1.856$ (0.175, 20) & $-1.840$ & $-1.868$ (0.169, 26) \\
10065.045 & 4.8349 & $-0.289$ & $-0.600$ (0.129, 31) & $-0.570$ & $-0.597$ (0.132, 29) \\
10114.020 & 2.7600 & \multicolumn{1}{c}{---} & \multicolumn{1}{c}{---} & $-3.760$ & $-3.711$ (0.147, 23) \\
10145.561 & 4.7955 & $-0.177$ & $-0.349$ (0.109, 47) & $-0.410$ & $-0.382$ (0.105, 48) \\
10155.162 & 2.1759 & $-4.226$ & $-4.611$ (0.162, 26) & $-4.360$ & $-4.518$ (0.091, 38) \\
10167.468 & 2.1979 & $-4.117$ & $-4.350$ (0.103, 41) & $-4.260$ & $-4.380$ (0.104, 39) \\
10195.105 & 2.7275 & $-3.580$ & $-3.765$ (0.109, 43) & $-3.630$ & $-3.781$ (0.109, 46) \\
10216.313 & 4.7331 & $-0.063$ & $-0.226$ (0.126, 49) & $-0.290$ & $-0.232$ (0.137, 50) \\
10218.408 & 3.0713 & $-2.760$ & $-3.014$ (0.101, 48) & $-2.930$ & $-3.044$ (0.093, 49) \\
10227.994 & 6.1189 & $-0.354$ & $-0.299$ (0.093, 38) & \multicolumn{1}{c}{---} & \multicolumn{1}{c}{---} \\
10265.217 & 2.2227 & $-4.537$ & $-4.750$ (0.096, 23) & \multicolumn{1}{c}{---} & \multicolumn{1}{c}{---} \\
10340.885 & 2.1979 & $-3.577$ & $-3.789$ (0.096, 48) & $-3.650$ & $-3.814$ (0.086, 49) \\
10347.965 & 5.3933 & $-0.551$ & $-0.839$ (0.058, 48) & $-0.820$ & $-0.894$ (0.051, 50) \\
10353.804 & 5.3933 & $-0.819$ & $-1.098$ (0.071, 48) & $-1.090$ & $-1.108$ (0.070, 50) \\
10395.794 & 2.1759 & $-3.393$ & $-3.455$ (0.076, 28) & $-3.420$ & $-3.570$ (0.118, 36) \\
10469.652 & 3.8835 & $-1.184$ & $-1.736$ (0.098, 39) & $-1.370$ & $-1.758$ (0.104, 37) \\
10532.234 & 3.9286 & $-1.480$ & $-1.918$ (0.076, 51) & $-1.760$ & $-1.960$ (0.066, 50) \\
10535.709 & 6.2057 & $-0.108$ & $-0.219$ (0.075, 50) & \multicolumn{1}{c}{---} & \multicolumn{1}{c}{---} \\
10577.139 & 3.3014 & $-3.136$ & $-3.452$ (0.123, 40) & $-3.280$ & $-3.519$ (0.137, 41) \\
10611.686 & 6.1692 & $ 0.021$ & $-0.090$ (0.066, 51) & $-0.090$ & $-0.110$ (0.065, 51) \\
10616.721 & 3.2671 & $-3.127$ & $-3.536$ (0.093, 22) & $-3.340$ & $-3.537$ (0.110, 45) \\
10674.070 & 6.1692 & $-0.466$ & $-0.586$ (0.196, 22) & \multicolumn{1}{c}{---} & \multicolumn{1}{c}{---} \\
10725.185 & 3.6398 & $-2.763$ & $-2.937$ (0.123, 23) & $-2.980$ & $-2.948$ (0.125, 22) \\
10753.004 & 3.9597 & $-1.845$ & $-2.217$ (0.097, 26) & $-2.140$ & $-2.214$ (0.094, 27) \\
10818.274 & 3.9597 & $-1.948$ & $-2.292$ (0.086, 47) & $-2.230$ & $-2.325$ (0.095, 41) \\
10849.460 & 5.5400 & \multicolumn{1}{c}{---} & \multicolumn{1}{c}{---} & $-0.730$ & $-0.828$ (0.083, 44) \\
10863.518 & 4.7331 & $-0.895$ & $-1.032$ (0.093, 36) & $-1.060$ & $-1.135$ (0.075, 49) \\
12053.082 & 4.5585 & $-1.543$ & $-1.767$ (0.136, 44) & $-1.750$ & $-1.803$ (0.128, 44) \\
12119.494 & 4.5931 & $-1.635$ & $-1.977$ (0.180, 20) & \multicolumn{1}{c}{---} & \multicolumn{1}{c}{---} \\
12190.098 & 3.6352 & $-2.330$ & $-2.748$ (0.131, 38) & $-2.750$ & $-2.823$ (0.124, 41) \\
12283.298 & 6.1692 & $-0.537$ & $-0.651$ (0.180, 20) & $-0.610$ & $-0.584$ (0.144, 22) \\
12342.916 & 4.6382 & $-1.463$ & $-1.703$ (0.122, 42) & $-1.680$ & $-1.690$ (0.135, 33) \\
12556.996 & 2.2786 & $-3.626$ & $-4.110$ (0.128, 37) & $-4.070$ & $-4.147$ (0.144, 32) \\
12638.703 & 4.5585 & $-0.783$ & $-1.153$ (0.165, 35) & $-1.000$ & $-1.164$ (0.161, 34) \\
12648.741 & 4.6070 & $-1.140$ & $-1.325$ (0.153, 33) & $-1.320$ & $-1.364$ (0.158, 35) \\
12789.470 & 5.0100 & \multicolumn{1}{c}{---} & \multicolumn{1}{c}{---} & $-1.920$ & $-1.704$ (0.148, 23) \\
12807.152 & 3.6398 & $-2.452$ & $-2.710$ (0.166, 20) & $-2.760$ & $-2.728$ (0.122, 26) \\
12879.766 & 2.2786 & $-3.458$ & $-3.686$ (0.115, 31) & $-3.610$ & $-3.714$ (0.103, 33) \\
13006.684 & 2.9904 & $-3.744$ & $-3.493$ (0.142, 20) & \multicolumn{1}{c}{---} & \multicolumn{1}{c}{---} \\
\end{longtable}
}

{\small
\begin{longtable}{ccrrrrr}
\caption{Calibration of $\log gf$ of lines other than $\mathrm{Fe\,I}$ \label{tab:calib_species_loggf}} \\
\hline\hline
 & \multicolumn{1}{c}{$\lambda_\mathrm{air}$} & \multicolumn{1}{c}{EP} & \multicolumn{2}{l}{$\log gf_\mathrm{VALD}$ (dex)} & \multicolumn{2}{l}{$\log gf_\mathrm{MB99}$ (dex)} \\
 & \multicolumn{1}{c}{(\AA)} & \multicolumn{1}{c}{(eV)} & \multicolumn{1}{c}{old} & \multicolumn{1}{c}{new (SD, $N$)} & \multicolumn{1}{c}{old} & \multicolumn{1}{c}{new (SD, $N$)} \\
\hline
\endfirsthead
\caption[]{Calibration of $\log gf$ of lines other than $\mathrm{Fe\,I}$---continued.} \\
\hline
 & \multicolumn{1}{c}{$\lambda_\mathrm{air}$} & \multicolumn{1}{c}{EP} & \multicolumn{2}{l}{$\log gf_\mathrm{VALD}$ (dex)} & \multicolumn{2}{l}{$\log gf_\mathrm{MB99}$ (dex)} \\
 & \multicolumn{1}{c}{(\AA)} & \multicolumn{1}{c}{(eV)} & \multicolumn{1}{c}{old} & \multicolumn{1}{c}{new (SD, $N$)} & \multicolumn{1}{c}{old} & \multicolumn{1}{c}{new (SD, $N$)} \\
\hline
\endhead
\hline
\endfoot
\ion{Si}{i} & 10068.329 &  6.099 & $-1.318$ & $-1.566$ (0.071, 37) & $-1.400$ & $-1.564$ (0.070, 39) \\
\ion{Si}{i} & 10288.944 &  4.920 & $-1.511$ & $-1.851$ (0.078, 51) & $-1.710$ & $-1.885$ (0.074, 50) \\
\ion{Si}{i} & 10301.410 &  6.100 & \multicolumn{1}{c}{---} & \multicolumn{1}{c}{---}  & $-1.830$ & $-1.946$ (0.121, 31) \\
\ion{Si}{i} & 10313.197 &  6.399 & $-0.886$ & $-1.460$ (0.125, 31) & \multicolumn{1}{c}{---} & \multicolumn{1}{c}{---}  \\
\ion{Si}{i} & 10407.037 &  6.616 & $-0.597$ & $-0.937$ (0.072, 48) & $-0.770$ & $-0.891$ (0.074, 48) \\
\ion{Si}{i} & 10414.913 &  6.619 & $-1.137$ & $-1.579$ (0.113, 45) & $-1.380$ & $-1.563$ (0.095, 46) \\
\ion{Si}{i} & 10582.160 &  6.223 & $-1.169$ & $-1.218$ (0.088, 21) & \multicolumn{1}{c}{---} & \multicolumn{1}{c}{---}  \\
\ion{Si}{i} & 10784.562 &  5.964 & $-0.839$ & $-0.838$ (0.088, 45) & $-0.720$ & $-0.886$ (0.076, 45) \\
\ion{Si}{i} & 10796.106 &  6.181 & $-1.266$ & $-1.628$ (0.104, 40) & $-1.490$ & $-1.561$ (0.123, 39) \\
\ion{Si}{i} & 10882.809 &  5.984 & $-0.815$ & $-0.812$ (0.085, 39) & $-0.620$ & $-0.789$ (0.077, 41) \\
\ion{Si}{i} & 12110.659 &  6.616 & $-0.136$ & $-0.411$ (0.147, 23) & \multicolumn{1}{c}{---} & \multicolumn{1}{c}{---}  \\
\ion{Si}{i} & 12175.733 &  6.619 & $-0.855$ & $-1.061$ (0.106, 33) & $-0.970$ & $-1.052$ (0.116, 39) \\
\ion{Si}{i} & 12178.339 &  6.269 & $-1.100$ & $-1.161$ (0.122, 45) & $-1.140$ & $-1.179$ (0.124, 45) \\
\ion{Si}{i} & 12390.154 &  5.082 & $-1.767$ & $-1.977$ (0.098, 47) & $-1.930$ & $-2.006$ (0.096, 48) \\
\ion{Si}{i} & 12395.832 &  4.954 & $-1.644$ & $-1.871$ (0.084, 47) & $-1.820$ & $-1.874$ (0.085, 48) \\
\ion{Si}{i} & 12583.924 &  6.616 & $-0.462$ & $-0.736$ (0.145, 37) & $-0.620$ & $-0.751$ (0.124, 36) \\
\ion{P}{i} & 9796.8280 &  6.985 & $ 0.270$ & $ 0.338$ (0.104, 38) & \multicolumn{1}{c}{---} & \multicolumn{1}{c}{---}  \\
\ion{P}{i} & 9903.6710 &  7.176 & $-0.300$ & $-0.322$ (0.142, 24) & \multicolumn{1}{c}{---} & \multicolumn{1}{c}{---}  \\
\ion{P}{i} & 10084.277 &  7.213 & $ 0.140$ & $ 0.202$ (0.129, 30) & $-0.070$ & $ 0.141$ (0.140, 36) \\
\ion{P}{i} & 10204.700 &  7.210 & \multicolumn{1}{c}{---} & \multicolumn{1}{c}{---}  & $-0.590$ & $-0.543$ (0.115, 24) \\
\ion{P}{i} & 10511.588 &  6.936 & $-0.130$ & $-0.192$ (0.116, 42) & $-0.220$ & $-0.195$ (0.110, 42) \\
\ion{P}{i} & 10529.524 &  6.954 & $ 0.240$ & $ 0.432$ (0.055, 46) & $ 0.140$ & $ 0.360$ (0.067, 46) \\
\ion{P}{i} & 10581.577 &  6.985 & $ 0.450$ & $ 0.660$ (0.132, 44) & $ 0.360$ & $ 0.678$ (0.121, 45) \\
\ion{P}{i} & 10596.903 &  6.936 & $-0.210$ & $-0.005$ (0.075, 49) & $-0.280$ & $-0.074$ (0.096, 49) \\
\ion{P}{i} & 10813.141 &  6.985 & $-0.410$ & $-0.395$ (0.128, 33) & $-0.440$ & $-0.409$ (0.153, 36) \\
\ion{S}{i} & 10635.970 &  8.580 & \multicolumn{1}{c}{---} & \multicolumn{1}{c}{---}  & $ 0.380$ & $ 0.478$ (0.116, 49) \\
\ion{S}{i} & 10821.180 &  0.000 & $-8.607$ & $-8.762$ (0.074, 32) & $-8.550$ & $-8.743$ (0.101, 39) \\
\ion{Ca}{i} & 10343.819 &  2.932 & $-0.300$ & $-0.446$ (0.106, 50) & $-0.400$ & $-0.439$ (0.105, 49) \\
\ion{Ca}{i} & 10516.140 &  4.740 & \multicolumn{1}{c}{---} & \multicolumn{1}{c}{---}  & $-0.520$ & $-0.765$ (0.160, 33) \\
\ion{Ca}{i} & 10791.450 &  4.740 & \multicolumn{1}{c}{---} & \multicolumn{1}{c}{---}  & $-0.680$ & $-0.740$ (0.132, 21) \\
\ion{Ca}{i} & 10838.970 &  4.878 & $ 0.238$ & $-0.141$ (0.113, 44) & $ 0.030$ & $-0.140$ (0.106, 43) \\
\ion{Ca}{i} & 10846.790 &  4.740 & \multicolumn{1}{c}{---} & \multicolumn{1}{c}{---}  & $-0.640$ & $-0.565$ (0.086, 39) \\
\ion{Ca}{i} & 11955.955 &  4.131 & $-0.849$ & $-1.065$ (0.180, 22) & $-0.910$ & $-1.060$ (0.183, 21) \\
\ion{Ca}{i} & 12105.840 &  4.550 & \multicolumn{1}{c}{---} & \multicolumn{1}{c}{---}  & $-0.540$ & $-0.726$ (0.168, 24) \\
\ion{Ca}{i} & 13033.554 &  4.441 & $-0.064$ & $-0.415$ (0.159, 32) & $-0.310$ & $-0.379$ (0.164, 34) \\
\ion{Ca}{i} & 13134.942 &  4.451 & $ 0.085$ & $-0.181$ (0.198, 22) & $-0.140$ & $-0.185$ (0.198, 22) \\
\ion{Ca}{ii} & 9854.7590 &  7.505 & $-0.205$ & $-0.277$ (0.134, 32) & \multicolumn{1}{c}{---} & \multicolumn{1}{c}{---}  \\
\ion{Fe}{ii} & 9956.3220 &  5.484 & $-2.985$ & $-3.017$ (0.130, 35) & \multicolumn{1}{c}{---} & \multicolumn{1}{c}{---}  \\
\ion{Fe}{ii} & 9997.5980 &  5.484 & $-1.867$ & $-1.721$ (0.164, 40) & \multicolumn{1}{c}{---} & \multicolumn{1}{c}{---}  \\
\ion{Fe}{ii} & 10173.515 &  5.511 & $-2.736$ & $-2.805$ (0.103, 43) & \multicolumn{1}{c}{---} & \multicolumn{1}{c}{---}  \\
\ion{Fe}{ii} & 10189.060 &  6.729 & $-2.141$ & $-2.225$ (0.127, 32) & \multicolumn{1}{c}{---} & \multicolumn{1}{c}{---}  \\
\ion{Fe}{ii} & 10245.556 &  6.730 & $-2.057$ & $-1.865$ (0.198, 28) & \multicolumn{1}{c}{---} & \multicolumn{1}{c}{---}  \\
\ion{Fe}{ii} & 10332.928 &  6.729 & $-1.968$ & $-1.965$ (0.100, 37) & \multicolumn{1}{c}{---} & \multicolumn{1}{c}{---}  \\
\ion{Fe}{ii} & 10366.167 &  6.724 & $-1.825$ & $-1.860$ (0.114, 47) & \multicolumn{1}{c}{---} & \multicolumn{1}{c}{---}  \\
\ion{Fe}{ii} & 10490.945 &  5.549 & $-2.985$ & $-2.915$ (0.101, 42) & \multicolumn{1}{c}{---} & \multicolumn{1}{c}{---}  \\
\ion{Fe}{ii} & 10501.503 &  5.548 & $-2.086$ & $-1.975$ (0.086, 50) & \multicolumn{1}{c}{---} & \multicolumn{1}{c}{---}  \\
\ion{Fe}{ii} & 10525.149 &  5.553 & $-2.958$ & $-2.938$ (0.092, 46) & \multicolumn{1}{c}{---} & \multicolumn{1}{c}{---}  \\
\ion{Fe}{ii} & 10862.652 &  5.589 & $-2.199$ & $-2.327$ (0.107, 26) & \multicolumn{1}{c}{---} & \multicolumn{1}{c}{---}  \\
\ion{Fe}{ii} & 10871.626 &  5.589 & $-3.098$ & $-2.974$ (0.111, 24) & \multicolumn{1}{c}{---} & \multicolumn{1}{c}{---}  \\
\ion{Zn}{i} & 13053.600 &  6.655 & $ 0.340$ & $ 0.329$ (0.176, 35) & $ 0.130$ & $ 0.320$ (0.163, 32) \\
\ion{Y}{ii} & 10105.520 &  1.721 & $-1.890$ & $-1.672$ (0.126, 34) & $-1.890$ & $-1.627$ (0.130, 35) \\
\ion{Y}{ii} & 10186.460 &  1.839 & \multicolumn{1}{c}{---} & \multicolumn{1}{c}{---}  & $-1.970$ & $-2.215$ (0.148, 25) \\
\ion{Y}{ii} & 10245.220 &  1.738 & $-1.820$ & $-2.167$ (0.191, 36) & $-1.910$ & $-2.246$ (0.230, 25) \\
\ion{Y}{ii} & 10329.700 &  1.748 & $-1.760$ & $-1.877$ (0.121, 24) & $-1.710$ & $-1.885$ (0.125, 34) \\
\ion{Y}{ii} & 10605.150 &  1.738 & $-1.960$ & $-1.844$ (0.118, 43) & $-1.890$ & $-1.844$ (0.119, 41) \\
\ion{Dy}{ii} & 10523.390 &  1.946 & $-0.450$ & $-0.357$ (0.176, 21) & \multicolumn{1}{c}{---} & \multicolumn{1}{c}{---}  \\
\end{longtable}
}

\section{Derived abundances for the calibrators and validators sample.}
\label{AppendixC}

{\small
\begin{longtable}{ccrrrr}
\caption{Derived microturbulence and \ion{Fe}{i} abundance\label{tab:xi_Fe}} \\
\hline\hline
 Cepheid & Date (Phase) & \multicolumn{1}{c}{$T_\mathrm{LDR}$} & \multicolumn{1}{c}{$\log g_\mathrm{rel}$} & \multicolumn{1}{c}{$\xi$} & \multicolumn{1}{c}{[Fe/H] (SD, $N$)} \\
  &  & \multicolumn{1}{c}{(K)} & \multicolumn{1}{c}{(dex)} & \multicolumn{1}{c}{(km\,s{$^{-1}$})} & \multicolumn{1}{c}{(dex)} \\
\hline
\endfirsthead
\caption[]{Derived microturbulence and \ion{Fe}{i} abundance---continued.} \\
\hline
 Cepheid & Date (Phase) & \multicolumn{1}{c}{$T_\mathrm{LDR}$} & \multicolumn{1}{c}{$\log g_\mathrm{rel}$} & \multicolumn{1}{c}{$\xi$} & \multicolumn{1}{c}{[Fe/H] (SD, $N$)} \\
  &  & \multicolumn{1}{c}{(K)} & \multicolumn{1}{c}{(dex)} & \multicolumn{1}{c}{(km\,s{$^{-1}$})} & \multicolumn{1}{c}{(dex)} \\
\hline
\endhead
\hline
\endfoot
\hline
\multicolumn{6}{c}{\textit{Calibrators}} \\ 
$\delta$ Cep & 20131202 (0.119) & 6159 & 2.08 & 2.40 & $ 0.10$ (0.14, 52) \\
             & 20131205 (0.672) & 5461 & 1.74 & 3.40 & $ 0.04$ (0.14, 53) \\
             & 20150806 (0.178) & 5975 & 1.99 & 3.00 & $ 0.14$ (0.17, 35) \\
             & 20150808 (0.556) & 5565 & 1.79 & 3.40 & $ 0.04$ (0.13, 59) \\
             & 20150815 (0.897) & 6306 & 2.14 & 3.60 & $ 0.01$ (0.17, 35) \\
             & 20151023 (0.711) & 5531 & 1.78 & 3.60 & $ 0.02$ (0.13, 61) \\
\hline
$\eta$ Aql   & 20140916 (0.743) & 5357 & 1.59 & 4.20 & $ 0.03$ (0.15, 47) \\
             & 20150806 (0.887) & 5887 & 1.85 & 3.00 & $ 0.14$ (0.17, 35) \\
             & 20150808 (0.174) & 5851 & 1.84 & 2.80 & $ 0.15$ (0.15, 51) \\
             & 20160321 (0.700) & 5304 & 1.56 & 4.20 & $-0.04$ (0.15, 42) \\
             & 20160326 (0.395) & 5675 & 1.75 & 3.00 & $ 0.06$ (0.16, 52) \\
             & 20160514 (0.216) & 5847 & 1.83 & 2.40 & $ 0.17$ (0.15, 44) \\
\hline
FF Aql       & 20150806 (0.594) & 5918 & 2.03 & 3.00 & $ 0.02$ (0.16, 37) \\
             & 20150807 (0.823) & 6155 & 2.14 & 3.40 & $-0.08$ (0.15, 43) \\
             & 20160317 (0.737) & 6300 & 2.20 & 3.40 & $-0.02$ (0.14, 52) \\
             & 20160321 (0.627) & 6156 & 2.14 & 5.20 & $-0.13$ (0.15, 46) \\
             & 20160419 (0.105) & 6258 & 2.18 & 3.20 & $ 0.08$ (0.13, 57) \\
\hline
RT Aur       & 20140123 (0.920) & 6206 & 2.22 & 3.20 & $-0.05$ (0.13, 57) \\
             & 20151028 (0.422) & 5712 & 1.99 & 2.20 & $ 0.07$ (0.13, 62) \\
             & 20160216 (0.149) & 6304 & 2.27 & 2.20 & $ 0.13$ (0.16, 38) \\
             & 20160228 (0.341) & 5852 & 2.06 & 1.80 & $ 0.10$ (0.12, 66) \\
             & 20160307 (0.492) & 5681 & 1.97 & 2.40 & $ 0.10$ (0.13, 61) \\
             & 20160315 (0.620) & 5667 & 1.97 & 3.00 & $ 0.02$ (0.12, 65) \\
             & 20160317 (0.155) & 6383 & 2.30 & 2.80 & $ 0.12$ (0.14, 54) \\
             & 20160321 (0.252) & 6069 & 2.16 & 2.40 & $ 0.07$ (0.13, 61) \\
             & 20160323 (0.776) & 5737 & 2.00 & 3.60 & $-0.01$ (0.12, 65) \\
\hline
SU Cas       & 20150815 (0.014) &  nan &  nan & 1.80 & $ 0.04$ (0.17, 34) \\
             & 20160312 (0.546) & 6104 & 2.39 & 2.20 & $-0.07$ (0.13, 57) \\
             & 20160315 (0.108) & 6341 & 2.50 & 1.80 & $ 0.01$ (0.15, 44) \\
             & 20160317 (0.150) & 6420 & 2.54 & 2.20 & $-0.04$ (0.14, 49) \\
\hline
SZ Tau       & 20140124 (0.870) & 6100 & 2.23 & 3.40 & $-0.07$ (0.15, 46) \\
             & 20160218 (0.639) & 5799 & 2.09 & 3.00 & $-0.06$ (0.13, 57) \\
             & 20160317 (0.502) & 5813 & 2.10 & 2.40 & $-0.02$ (0.13, 62) \\
             & 20160321 (0.772) & 6115 & 2.24 & 2.40 & $-0.05$ (0.16, 48) \\
\hline
X Cyg        & 20140830 (0.606) & 4727 & 0.96 & 4.00 & $ 0.07$ (0.14, 49) \\
             & 20140915 (0.582) & 4717 & 0.95 & 3.60 & $ 0.11$ (0.14, 54) \\
             & 20140918 (0.760) & 5102 & 1.17 & 5.80 & $ 0.02$ (0.18, 58) \\
             & 20141015 (0.408) & 4912 & 1.07 & 3.80 & $ 0.08$ (0.13, 60) \\
             & 20150725 (0.685) & 4852 & 1.03 & 4.60 & $ 0.12$ (0.15, 46) \\
             & 20151026 (0.353) & 5059 & 1.15 & 2.40 & $ 0.07$ (0.14, 49) \\
             & 20160317 (0.103) & 5629 & 1.45 & 3.20 & $ 0.04$ (0.13, 62) \\
             & 20160504 (0.030) & 6007 & 1.63 & 3.40 & $ 0.10$ (0.13, 57) \\
             & 20160518 (0.880) & 5384 & 1.32 & 3.60 & $ 0.26$ (0.13, 59) \\
\hline
$\zeta$ Gem  & 20130222 (0.534) & 5246 & 1.41 & 2.80 & $ 0.01$ (0.12, 65) \\
             & 20130223 (0.645) & 5443 & 1.52 & 4.60 & $-0.06$ (0.13, 56) \\
             & 20130227 (0.028) & 5803 & 1.70 & 3.20 & $-0.13$ (0.14, 55) \\
             & 20131129 (0.133) & 5637 & 1.61 & 3.60 & $-0.04$ (0.12, 69) \\
             & 20160217 (0.923) & 5711 & 1.65 & 2.60 & $ 0.00$ (0.15, 47) \\
             & 20160307 (0.791) & 5707 & 1.65 & 3.40 & $ 0.03$ (0.14, 53) \\
             & 20160501 (0.200) & 5488 & 1.54 & 3.60 & $ 0.04$ (0.13, 58) \\
             & 20171204 (0.626) & 5509 & 1.55 & 4.60 & $-0.02$ (0.15, 44) \\

\hline
DL Cas       & 20150731 (0.743) & 5422 & 1.59 & 3.20 & $-0.10$ (0.15, 43) \\
             & 20150807 (0.604) & 5310 & 1.53 & 3.80 & $-0.12$ (0.14, 58) \\
             & 20151023 (0.219) & 5856 & 1.80 & 3.20 & $-0.00$ (0.12, 65) \\
\hline
S Sge        & 20151026 (0.940) & 6097 & 1.90 & 3.00 & $ 0.03$ (0.16, 40) \\
             & 20160321 (0.520) & 5343 & 1.53 & 3.40 & $-0.03$ (0.13, 59) \\
             & 20160326 (0.119) & 5997 & 1.85 & 3.60 & $ 0.03$ (0.14, 52) \\
\hline
T Vul        & 20150808 (0.138) & 6131 & 2.13 & 2.80 & $-0.01$ (0.15, 47) \\
             & 20160514 (0.293) & 5799 & 1.97 & 2.20 & $-0.04$ (0.14, 52) \\
\hline
\end{longtable}
}

{\footnotesize
\begin{longtable}{ccrrr}
\caption{Abundances other than $\mathrm{Fe\,I}$ (Part 1) \label{tab:species_results1}} \\
\hline\hline
 Cepheid & Date  & \multicolumn{1}{c}{$\mathrm{Si\,I}$} & \multicolumn{1}{c}{$\mathrm{P\,I}$} & \multicolumn{1}{c}{$\mathrm{S\,I}$}\\
\hline
\endfirsthead
\caption[]{Abundances other than  $\mathrm{Fe\,I}$ (Part 1 ---continued)} \\
\hline
 Cepheid & Date  & \multicolumn{1}{c}{$\mathrm{Si\,I}$} & \multicolumn{1}{c}{$\mathrm{P\,I}$} & \multicolumn{1}{c}{$\mathrm{S\,I}$}\\
\hline
\endhead
\hline
\endfoot
\hline
$\delta$ Cep & 20131202 & $0.33$ (0.08, 19) & $0.11$ (0.05, 10) & $0.32$ (0.05, 3) \\ 
 & 20131205 & $0.21$ (0.04, 22) & $-0.01$ (0.05, 9) & $0.20$ (0.06, 3) \\ 
 & 20150806 & $0.29$ (0.12, 12) & $0.10$ (0.04, 9) & $0.25$ (0.12, 1) \\ 
 & 20150808 & $0.24$ (0.05, 16) & $0.00$ (0.05, 11) & $0.27$ (0.08, 3) \\ 
 & 20150815 & $0.25$ (0.10, 21) & $-0.01$ (0.04, 12) & $-0.00$ (0.27, 3) \\ 
 & 20151023 & $0.19$ (0.04, 24) & $-0.03$ (0.03, 12) & $0.15$ (0.05, 3) \\ 
\hline
$\eta$ Aql & 20140916 & $0.18$ (0.18, 13) & $0.08$ (0.20, 9) & $0.52$ (0.18, 2) \\ 
 & 20150806 & $0.33$ (0.13, 19) & $0.19$ (0.16, 11) & $0.14$ (0.11, 2) \\ 
 & 20150808 & $0.30$ (0.11, 26) & $0.12$ (0.11, 14) & $0.35$ (0.10, 3) \\ 
 & 20160321 & $0.12$ (0.08, 22) & $0.02$ (0.12, 11) & $0.27$ (0.17, 3) \\ 
 & 20160326 & $0.28$ (0.08, 26) & $0.07$ (0.07, 12) & $0.34$ (0.12, 3) \\ 
 & 20160514 & $0.32$ (0.06, 22) & $0.15$ (0.08, 10) & $0.39$ (0.05, 3) \\ 
\hline
FF Aql & 20150806 & $0.13$ (0.11, 15) & $0.05$ (0.10, 9) & $0.29$ (0.12, 1) \\ 
 & 20150807 & $0.19$ (0.12, 22) & $0.06$ (0.07, 8) & $0.36$ (0.12, 1) \\ 
 & 20160317 & $0.21$ (0.07, 25) & $0.06$ (0.03, 11) & $0.16$ (0.23, 2) \\ 
 & 20160321 & $0.10$ (0.09, 23) & $-0.01$ (0.08, 13) & $0.17$ (0.08, 3) \\ 
 & 20160419 & $0.24$ (0.07, 25) & $0.05$ (0.06, 13) & $0.29$ (0.05, 3) \\ 
\hline
RT Aur & 20140123 & $0.19$ (0.07, 21) & $0.03$ (0.04, 12) & $0.36$ (0.12, 1) \\ 
 & 20151028 & $0.27$ (0.06, 26) & $0.08$ (0.04, 12) & $0.31$ (0.06, 3) \\ 
 & 20160216 & $0.30$ (0.11, 27) & $0.09$ (0.05, 14) & $0.44$ (0.12, 1) \\ 
 & 20160228 & $0.31$ (0.06, 25) & $0.07$ (0.07, 15) & $0.29$ (0.05, 3) \\ 
 & 20160307 & $0.21$ (0.07, 25) & $0.05$ (0.06, 12) & $0.22$ (0.05, 3) \\ 
 & 20160315 & $0.20$ (0.07, 22) & $0.03$ (0.05, 11) & $0.25$ (0.09, 3) \\ 
 & 20160317 & $0.27$ (0.07, 21) & $0.07$ (0.06, 12) & $0.36$ (0.11, 2) \\ 
 & 20160321 & $0.27$ (0.06, 27) & $0.03$ (0.08, 15) & $0.22$ (0.12, 1) \\ 
 & 20160323 & $0.20$ (0.07, 18) & $0.02$ (0.06, 9) & $0.26$ (0.05, 3) \\ 
\hline
SU Cas & 20150815 & $0.27$ (0.10, 16) & $0.09$ (0.13, 12) & $0.45$ (0.12, 1) \\ 
 & 20160312 & $0.20$ (0.09, 26) & $0.04$ (0.05, 14) & $0.21$ (0.08, 2) \\ 
 & 20160315 & $0.21$ (0.05, 21) & $0.04$ (0.09, 11) & $0.36$ (0.12, 1) \\ 
 & 20160317 & $0.18$ (0.06, 21) & $0.04$ (0.04, 14) & $0.26$ (0.12, 1) \\ 
\hline
SZ Tau & 20140124 & $0.10$ (0.09, 20) & $-0.04$ (0.06, 12) & $0.20$ (0.06, 2) \\ 
 & 20160218 & $0.16$ (0.06, 23) & $0.00$ (0.05, 14) & $0.17$ (0.08, 2) \\ 
 & 20160317 & $0.21$ (0.04, 24) & $0.03$ (0.05, 15) & $0.17$ (0.08, 2) \\ 
 & 20160321 & $0.18$ (0.13, 21) & $0.02$ (0.08, 12) & $0.34$ (0.12, 1) \\ 
\hline
X Cyg & 20140830 & $0.17$ (0.16, 13) & $-0.09$ (0.11, 3) & $0.29$ (0.35, 3) \\ 
 & 20140915 & $0.21$ (0.17, 14) & $-0.18$ (0.29, 4) & $0.35$ (0.24, 3) \\ 
 & 20140918 & $0.21$ (0.20, 17) & $-0.13$ (0.38, 4) & $0.15$ (0.08, 3) \\ 
 & 20141015 & $0.10$ (0.11, 15) & $0.05$ (0.13, 6) & $0.39$ (0.16, 3) \\ 
 & 20150725 & $0.25$ (0.18, 8) & $-0.10$ (0.31, 5) & $0.25$ (0.27, 3) \\ 
 & 20151026 & $0.26$ (0.07, 14) & $0.17$ (0.09, 12) & $0.34$ (0.07, 2) \\ 
 & 20160317 & $0.28$ (0.10, 20) & $0.06$ (0.05, 11) & $0.19$ (0.09, 3) \\ 
 & 20160504 & $0.29$ (0.13, 26) & $0.03$ (0.12, 14) & $0.37$ (0.10, 3) \\ 
 & 20160518 & $0.32$ (0.12, 22) & $-0.06$ (0.12, 8) & $-0.05$ (0.12, 1) \\ 
\hline
$\zeta$ Gem & 20130222 & $0.20$ (0.09, 22) & $-0.06$ (0.10, 11) & $0.12$ (0.12, 1) \\ 
 & 20130223 & $0.10$ (0.09, 21) & $-0.19$ (0.10, 10) & $0.18$ (0.20, 3) \\ 
 & 20130227 & $0.12$ (0.05, 25) & $-0.03$ (0.06, 11) & $0.15$ (0.09, 2) \\ 
 & 20131129 & $0.13$ (0.04, 26) & $-0.08$ (0.04, 15) & $0.15$ (0.06, 3) \\ 
 & 20160217 & $0.24$ (0.06, 23) & $0.02$ (0.04, 13) & $0.28$ (0.06, 3) \\ 
 & 20160307 & $0.21$ (0.03, 22) & $-0.05$ (0.03, 11) & $0.14$ (0.12, 1) \\ 
 & 20160501 & $0.17$ (0.05, 28) & $-0.01$ (0.06, 15) & $0.18$ (0.11, 3) \\ 
 & 20171204 & $0.14$ (0.06, 21) & $-0.09$ (0.03, 6) & $0.16$ (0.11, 3) \\ 
\hline
DL Cas & 20150731 & $0.23$ (0.11, 16) & $-0.09$ (0.18, 6) & $0.41$ (0.20, 2) \\ 
 & 20150807 & $0.21$ (0.24, 19) & $0.09$ (0.16, 7) & $0.16$ (0.06, 3) \\ 
 & 20151023 & $0.19$ (0.08, 24) & $-0.08$ (0.11, 9) & $0.39$ (0.05, 3) \\ 
\hline
S Sge & 20151026 & $0.25$ (0.08, 21) & $0.07$ (0.04, 10) & $0.42$ (0.12, 1) \\ 
 & 20160321 & $0.24$ (0.06, 18) & $0.03$ (0.04, 11) & $0.21$ (0.05, 3) \\ 
 & 20160326 & $0.24$ (0.07, 23) & $0.00$ (0.04, 14) & $0.23$ (0.05, 3) \\ 
\hline
T Vul & 20150808 & $0.23$ (0.07, 19) & $-0.03$ (0.07, 13) & $0.17$ (0.12, 1) \\ 
 & 20160514 & $0.18$ (0.09, 25) & $0.02$ (0.07, 13) & $0.24$ (0.05, 3) \\ 
\hline
\end{longtable}
\begin{longtable}{ccrrr}
\caption{Abundances other than $\mathrm{Fe\,I}$ (Part 2) \label{tab:species_results2}} \\
\hline\hline
 Cepheid & Date  & \multicolumn{1}{c}{$\mathrm{Ca\,I}$} & \multicolumn{1}{c}{$\mathrm{P\,I}$} & \multicolumn{1}{c}{$\mathrm{Fe\,II}$}\\
\hline
\endfirsthead
\caption[]{Abundances other than $\mathrm{Fe\,I}$ (Part 2 ---continued)} \\
\hline
 Cepheid & Date  & \multicolumn{1}{c}{$\mathrm{Ca\,I}$} & \multicolumn{1}{c}{$\mathrm{Ca\,II}$} & \multicolumn{1}{c}{$\mathrm{Fe\,II}$}\\
\hline
\endhead
\hline
\endfoot
\hline
$\delta$ Cep & 20131202 & $0.15$ (0.10, 8) & $0.60$ (0.13, 1) & $0.12$ (0.10, 9) \\ 
 & 20131205 & $0.21$ (0.05, 7) & $0.32$ (0.13, 1) & $0.03$ (0.08, 6) \\ 
 & 20150806 & \multicolumn{1}{c}{---}  & $0.70$ (0.13, 1) & $0.12$ (0.18, 8) \\ 
 & 20150808 & $0.22$ (0.08, 5) & \multicolumn{1}{c}{---}  & $0.01$ (0.06, 3) \\ 
 & 20150815 & $0.09$ (0.08, 4) & $0.61$ (0.13, 1) & $0.08$ (0.11, 8) \\ 
 & 20151023 & $0.20$ (0.07, 12) & $0.25$ (0.13, 1) & $0.02$ (0.06, 9) \\ 
\hline
$\eta$ Aql & 20140916 & $0.26$ (0.24, 7) & $0.45$ (0.13, 1) & $0.15$ (0.29, 7) \\ 
 & 20150806 & $0.33$ (0.11, 4) & \multicolumn{1}{c}{---}  & $0.10$ (0.14, 10) \\ 
 & 20150808 & $0.27$ (0.16, 7) & $0.49$ (0.13, 1) & $0.07$ (0.09, 10) \\ 
 & 20160321 & $0.18$ (0.14, 7) & \multicolumn{1}{c}{---}  & $-0.05$ (0.09, 7) \\ 
 & 20160326 & $0.25$ (0.10, 6) & $0.38$ (0.13, 1) & $-0.01$ (0.25, 8) \\ 
 & 20160514 & $0.18$ (0.11, 2) & $0.77$ (0.13, 1) & $0.11$ (0.16, 9) \\ 
\hline
FF Aql & 20150806 & $0.21$ (0.38, 3) & $0.69$ (0.13, 1) & $0.07$ (0.19, 6) \\ 
 & 20150807 & $0.31$ (0.20, 6) & $0.38$ (0.13, 1) & $0.06$ (0.10, 11) \\ 
 & 20160317 & $0.12$ (0.04, 6) & \multicolumn{1}{c}{---}  & $0.06$ (0.10, 12) \\ 
 & 20160321 & $0.05$ (0.17, 8) & \multicolumn{1}{c}{---}  & $-0.04$ (0.11, 11) \\ 
 & 20160419 & $0.18$ (0.06, 9) & \multicolumn{1}{c}{---}  & $0.03$ (0.09, 11) \\ 
\hline
RT Aur & 20140123 & $0.08$ (0.12, 11) & $0.52$ (0.13, 1) & $0.07$ (0.08, 11) \\ 
 & 20151028 & $0.26$ (0.09, 7) & $0.52$ (0.13, 1) & $0.10$ (0.06, 10) \\ 
 & 20160216 & $0.42$ (0.13, 8) & $0.66$ (0.13, 1) & $0.16$ (0.09, 10) \\ 
 & 20160228 & $0.22$ (0.04, 11) & $0.47$ (0.13, 1) & $0.05$ (0.07, 11) \\ 
 & 20160307 & $0.15$ (0.13, 7) & \multicolumn{1}{c}{---}  & $0.07$ (0.09, 11) \\ 
 & 20160315 & $0.19$ (0.08, 8) & \multicolumn{1}{c}{---}  & $0.01$ (0.06, 9) \\ 
 & 20160317 & $0.29$ (0.12, 12) & $0.37$ (0.13, 1) & $0.05$ (0.05, 11) \\ 
 & 20160321 & $0.21$ (0.07, 11) & \multicolumn{1}{c}{---}  & $-0.03$ (0.09, 11) \\ 
 & 20160323 & $0.28$ (0.12, 7) & $0.41$ (0.13, 1) & $0.08$ (0.06, 10) \\ 
\hline
SU Cas & 20150815 & $0.30$ (0.05, 5) & $0.68$ (0.13, 1) & $0.18$ (0.23, 8) \\ 
 & 20160312 & $0.12$ (0.07, 13) & $0.47$ (0.13, 1) & $0.00$ (0.09, 10) \\ 
 & 20160315 & $0.15$ (0.06, 7) & $0.81$ (0.13, 1) & $0.15$ (0.11, 10) \\ 
 & 20160317 & $0.08$ (0.12, 10) & $0.54$ (0.13, 1) & $0.03$ (0.04, 11) \\ 
\hline
SZ Tau & 20140124 & $-0.11$ (0.09, 7) & $0.33$ (0.13, 1) & $-0.00$ (0.08, 10) \\ 
 & 20160218 & $0.08$ (0.07, 10) & $0.34$ (0.13, 1) & $-0.01$ (0.08, 10) \\ 
 & 20160317 & $0.12$ (0.04, 10) & $0.46$ (0.13, 1) & $0.07$ (0.07, 7) \\ 
 & 20160321 & $0.14$ (0.17, 9) & $0.49$ (0.13, 1) & $0.05$ (0.09, 9) \\ 
\hline
X Cyg & 20140830 & $0.21$ (0.06, 8) & \multicolumn{1}{c}{---}  & $0.15$ (0.45, 3) \\ 
 & 20140915 & $0.32$ (0.19, 9) & $0.50$ (0.13, 1) & $0.01$ (0.28, 5) \\ 
 & 20140918 & $0.20$ (0.26, 4) & \multicolumn{1}{c}{---}  & $-0.07$ (0.18, 5) \\ 
 & 20141015 & $0.35$ (0.10, 8) & \multicolumn{1}{c}{---}  & $0.02$ (0.18, 5) \\ 
 & 20150725 & $0.13$ (0.22, 7) & \multicolumn{1}{c}{---}  & $-0.13$ (0.23, 3) \\ 
 & 20151026 & $0.29$ (0.07, 7) & $0.78$ (0.13, 1) & $0.22$ (0.15, 9) \\ 
 & 20160317 & $0.24$ (0.08, 3) & $0.45$ (0.13, 1) & $0.09$ (0.10, 8) \\ 
 & 20160504 & $0.25$ (0.07, 6) & $0.35$ (0.13, 1) & $0.12$ (0.07, 11) \\ 
 & 20160518 & $0.29$ (0.10, 6) & $0.08$ (0.13, 1) & $-0.05$ (0.13, 8) \\ 
\hline
$\zeta$ Gem & 20130222 & $0.14$ (0.09, 9) & \multicolumn{1}{c}{---}  & $-0.08$ (0.07, 5) \\ 
 & 20130223 & $0.10$ (0.12, 9) & \multicolumn{1}{c}{---}  & $-0.21$ (0.07, 7) \\ 
 & 20130227 & $0.12$ (0.09, 8) & $0.44$ (0.13, 1) & $-0.07$ (0.05, 7) \\ 
 & 20131129 & $0.10$ (0.04, 9) & $0.16$ (0.13, 1) & $-0.08$ (0.07, 11) \\ 
 & 20160217 & $0.13$ (0.04, 8) & $0.61$ (0.13, 1) & $0.03$ (0.06, 10) \\ 
 & 20160307 & $0.15$ (0.05, 6) & $0.36$ (0.13, 1) & $-0.06$ (0.06, 9) \\ 
 & 20160501 & $0.15$ (0.06, 10) & $0.17$ (0.13, 1) & $-0.07$ (0.07, 10) \\ 
 & 20171204 & $0.17$ (0.05, 5) & $0.09$ (0.13, 1) & $-0.19$ (0.16, 7) \\ 
\hline
DL Cas & 20150731 & $0.14$ (0.12, 4) & \multicolumn{1}{c}{---}  & $0.02$ (0.09, 6) \\ 
 & 20150807 & $0.19$ (0.14, 4) & \multicolumn{1}{c}{---}  & $0.03$ (0.22, 6) \\ 
 & 20151023 & $0.22$ (0.07, 5) & $0.20$ (0.13, 1) & $-0.01$ (0.10, 6) \\ 
\hline
S Sge & 20151026 & $0.21$ (0.05, 4) & $0.65$ (0.13, 1) & $0.11$ (0.08, 9) \\ 
 & 20160321 & $0.20$ (0.07, 7) & $0.25$ (0.13, 1) & $0.07$ (0.09, 6) \\ 
 & 20160326 & $0.14$ (0.05, 4) & $0.26$ (0.13, 1) & $0.06$ (0.04, 9) \\ 
\hline
T Vul & 20150808 & $0.12$ (0.07, 5) & \multicolumn{1}{c}{---}  & $-0.03$ (0.08, 8) \\ 
 & 20160514 & $0.08$ (0.12, 8) & $0.28$ (0.13, 1) & $-0.02$ (0.13, 9) \\ 
\hline
\end{longtable}
\begin{longtable}{ccrrr}
\caption{Abundances other than $\mathrm{Fe\,I}$ (Part 3) \label{tab:species_results3}} \\
\hline\hline
 Cepheid & Date  & \multicolumn{1}{c}{$\mathrm{Zn\,I}$} & \multicolumn{1}{c}{$\mathrm{Y\,II}$} & \multicolumn{1}{c}{$\mathrm{Dy\,II}$}\\
\hline
\endfirsthead
\caption[]{Abundances other than $\mathrm{Fe\,I}$ (Part 3 ---continued)} \\
\hline
 Cepheid & Date  & \multicolumn{1}{c}{$\mathrm{Zn\,I}$} & \multicolumn{1}{c}{$\mathrm{Y\,II}$} & \multicolumn{1}{c}{$\mathrm{Dy\,II}$}\\
\hline
\endhead
\hline
\endfoot
\hline
$\delta$ Cep & 20131202 & \multicolumn{1}{c}{---}  & $0.34$ (0.12, 4) & \multicolumn{1}{c}{---}  \\ 
 & 20131205 & $0.20$ (0.12, 2) & $0.20$ (0.10, 7) & \multicolumn{1}{c}{---}  \\ 
 & 20150806 & \multicolumn{1}{c}{---}  & $0.42$ (0.07, 8) & \multicolumn{1}{c}{---}  \\ 
 & 20150808 & \multicolumn{1}{c}{---}  & $0.33$ (0.07, 7) & \multicolumn{1}{c}{---}  \\ 
 & 20150815 & \multicolumn{1}{c}{---}  & $0.36$ (0.07, 3) & $0.02$ (0.18, 1) \\ 
 & 20151023 & $0.01$ (0.12, 2) & $0.27$ (0.07, 5) & $0.02$ (0.18, 1) \\ 
\hline
$\eta$ Aql & 20140916 & $-0.30$ (0.12, 2) & $0.37$ (0.10, 4) & \multicolumn{1}{c}{---}  \\ 
 & 20150806 & \multicolumn{1}{c}{---}  & $0.23$ (0.09, 2) & \multicolumn{1}{c}{---}  \\ 
 & 20150808 & \multicolumn{1}{c}{---}  & $0.34$ (0.08, 6) & $-0.13$ (0.18, 1) \\ 
 & 20160321 & $0.40$ (0.12, 2) & $0.33$ (0.20, 7) & $-0.35$ (0.18, 1) \\ 
 & 20160326 & $0.42$ (0.12, 2) & $0.35$ (0.07, 4) & \multicolumn{1}{c}{---}  \\ 
 & 20160514 & $-0.04$ (0.12, 2) & $0.40$ (0.09, 5) & \multicolumn{1}{c}{---}  \\ 
\hline
FF Aql & 20150806 & \multicolumn{1}{c}{---}  & $0.16$ (0.20, 7) & \multicolumn{1}{c}{---}  \\ 
 & 20150807 & \multicolumn{1}{c}{---}  & $0.09$ (0.09, 4) & $0.10$ (0.18, 1) \\ 
 & 20160317 & \multicolumn{1}{c}{---}  & $0.20$ (0.15, 6) & \multicolumn{1}{c}{---}  \\ 
 & 20160321 & \multicolumn{1}{c}{---}  & $0.15$ (0.05, 9) & \multicolumn{1}{c}{---}  \\ 
 & 20160419 & \multicolumn{1}{c}{---}  & $0.35$ (0.12, 7) & $0.01$ (0.18, 1) \\ 
\hline
RT Aur & 20140123 & $0.10$ (0.12, 2) & $0.32$ (0.07, 6) & \multicolumn{1}{c}{---}  \\ 
 & 20151028 & $0.14$ (0.12, 2) & $0.33$ (0.08, 6) & $0.23$ (0.18, 1) \\ 
 & 20160216 & $0.25$ (0.12, 2) & $0.41$ (0.12, 4) & \multicolumn{1}{c}{---}  \\ 
 & 20160228 & $0.07$ (0.12, 2) & $0.29$ (0.08, 6) & \multicolumn{1}{c}{---}  \\ 
 & 20160307 & $0.11$ (0.12, 2) & $0.31$ (0.06, 7) & \multicolumn{1}{c}{---}  \\ 
 & 20160315 & $0.08$ (0.12, 2) & $0.22$ (0.05, 6) & $-0.06$ (0.18, 1) \\ 
 & 20160317 & $0.32$ (0.12, 2) & $0.33$ (0.07, 5) & $0.12$ (0.18, 1) \\ 
 & 20160321 & $0.18$ (0.12, 2) & $0.31$ (0.05, 7) & $0.13$ (0.18, 1) \\ 
 & 20160323 & $-0.08$ (0.12, 2) & $0.21$ (0.17, 6) & \multicolumn{1}{c}{---}  \\ 
\hline
SU Cas & 20150815 & \multicolumn{1}{c}{---}  & $0.47$ (0.07, 4) & \multicolumn{1}{c}{---}  \\ 
 & 20160312 & $-0.05$ (0.12, 2) & $0.23$ (0.10, 5) & \multicolumn{1}{c}{---}  \\ 
 & 20160315 & $-0.14$ (0.20, 2) & $0.28$ (0.08, 2) & \multicolumn{1}{c}{---}  \\ 
 & 20160317 & $-0.10$ (0.18, 1) & $0.25$ (0.08, 2) & \multicolumn{1}{c}{---}  \\ 
\hline
SZ Tau & 20140124 & $-0.19$ (0.12, 2) & $0.21$ (0.11, 3) & \multicolumn{1}{c}{---}  \\ 
 & 20160218 & $-0.01$ (0.12, 2) & $0.13$ (0.08, 6) & \multicolumn{1}{c}{---}  \\ 
 & 20160317 & $0.05$ (0.12, 2) & $0.28$ (0.07, 5) & \multicolumn{1}{c}{---}  \\ 
 & 20160321 & $0.07$ (0.18, 1) & $0.30$ (0.11, 5) & $0.42$ (0.18, 1) \\ 
\hline
X Cyg & 20140830 & $0.08$ (0.12, 2) & $0.31$ (0.16, 8) & \multicolumn{1}{c}{---}  \\ 
 & 20140915 & $-0.10$ (0.16, 1) & $0.37$ (0.12, 6) & \multicolumn{1}{c}{---}  \\ 
 & 20140918 & $-0.51$ (0.18, 1) & $0.56$ (0.14, 4) & \multicolumn{1}{c}{---}  \\ 
 & 20141015 & $0.13$ (0.12, 2) & $0.55$ (0.13, 8) & \multicolumn{1}{c}{---}  \\ 
 & 20150725 & $0.04$ (0.12, 2) & $0.19$ (0.06, 4) & $-0.21$ (0.18, 1) \\ 
 & 20151026 & $0.07$ (0.12, 2) & $0.49$ (0.10, 7) & $0.24$ (0.18, 1) \\ 
 & 20160317 & $0.09$ (0.12, 2) & $0.36$ (0.06, 4) & $-0.01$ (0.18, 1) \\ 
 & 20160504 & $0.20$ (0.16, 1) & $0.32$ (0.06, 6) & \multicolumn{1}{c}{---}  \\ 
 & 20160518 & \multicolumn{1}{c}{---}  & $0.50$ (0.08, 3) & \multicolumn{1}{c}{---}  \\ 
\hline
$\zeta$ Gem & 20130222 & $-0.11$ (0.18, 1) & $0.25$ (0.10, 5) & \multicolumn{1}{c}{---}  \\ 
 & 20130223 & $-0.02$ (0.12, 2) & $0.08$ (0.06, 4) & $-0.14$ (0.18, 1) \\ 
 & 20130227 & $-0.02$ (0.12, 2) & $0.13$ (0.08, 6) & \multicolumn{1}{c}{---}  \\ 
 & 20131129 & $0.10$ (0.12, 2) & $0.20$ (0.06, 5) & $-0.01$ (0.18, 1) \\ 
 & 20160217 & $0.06$ (0.12, 2) & $0.31$ (0.06, 5) & $0.13$ (0.18, 1) \\ 
 & 20160307 & $0.03$ (0.12, 2) & $0.28$ (0.10, 2) & $-0.05$ (0.18, 1) \\ 
 & 20160501 & $0.21$ (0.12, 2) & $0.26$ (0.07, 6) & $0.08$ (0.18, 1) \\ 
 & 20171204 & \multicolumn{1}{c}{---}  & $0.21$ (0.14, 7) & $0.09$ (0.18, 1) \\ 
\hline
DL Cas & 20150731 & \multicolumn{1}{c}{---}  & $0.07$ (0.40, 4) & \multicolumn{1}{c}{---}  \\ 
 & 20150807 & \multicolumn{1}{c}{---}  & $0.28$ (0.17, 3) & \multicolumn{1}{c}{---}  \\ 
 & 20151023 & $0.10$ (0.12, 2) & $0.35$ (0.15, 2) & \multicolumn{1}{c}{---}  \\ 
\hline
S Sge & 20151026 & \multicolumn{1}{c}{---}  & $0.29$ (0.09, 2) & $-0.03$ (0.18, 1) \\ 
 & 20160321 & $-0.01$ (0.12, 2) & $0.34$ (0.09, 6) & $-0.04$ (0.18, 1) \\ 
 & 20160326 & $0.18$ (0.12, 2) & $0.31$ (0.08, 4) & \multicolumn{1}{c}{---}  \\ 
\hline
T Vul & 20150808 & \multicolumn{1}{c}{---}  & $0.27$ (0.07, 5) & $-0.14$ (0.18, 1) \\ 
 & 20160514 & $-0.08$ (0.18, 1) & $0.11$ (0.07, 5) & $0.04$ (0.18, 1) \\ 
\hline 
\end{longtable}
}

\bibliography{loggf_cal}{}
\bibliographystyle{aasjournal}



\end{document}